\documentclass[twocolumn,floats,floatfix,showpacs,amssymb,prd,superscriptaddress,nofootinbib]{revtex4-2}

\usepackage{amssymb,amsmath,verbatim,mathtools,needspace,enumitem,etoolbox,graphicx,physics,microtype,afterpage,xspace,tabularx,lmodern,multirow,bm}
\usepackage{dcolumn}
\usepackage{multirow}
\usepackage{float}
\usepackage{booktabs}
\usepackage{gensymb}
\usepackage{lipsum}
\usepackage[dvipsnames, usenames]{xcolor}
\definecolor{linkcolor}{rgb}{0.0,0.3,0.5}
\usepackage[unicode, colorlinks=true, linkcolor=linkcolor, citecolor=linkcolor, filecolor=linkcolor, urlcolor=linkcolor, linktocpage, breaklinks]{hyperref}
\usepackage[all]{hypcap}
\usepackage[T1]{fontenc}
\usepackage[utf8]{inputenc}
\usepackage[usenames,dvipsnames]{xcolor}
\definecolor{TurkishBlue}{HTML}{144893}
\definecolor{TurkishBlue2}{HTML}{1985B6}
\hypersetup{colorlinks=true,citecolor=TurkishBlue2,linkcolor=TurkishBlue2,urlcolor=TurkishBlue2}

\definecolor{romared}{RGB}{142,0,28}
\usepackage{aas_macros}
\usepackage{makecell}
\usepackage{soul}
\usepackage[nolist,nohyperlinks]{acronym}

\usepackage{array}

\usepackage{bm}
\renewcommand{\vec}{\bm}

\usepackage [english]{babel}
\usepackage [autostyle, english = american]{csquotes}
\MakeOuterQuote{"}

\newcommand{\approptoinn}[2]{\mathrel{\vcenter{
  \offinterlineskip\halign{\hfil$##$\cr
    #1\propto\cr\noalign{\kern2pt}#1\sim\cr\noalign{\kern-2pt}}}}}

\usepackage[normalem]{ulem}
\definecolor{mypink}{HTML}{C34D85}

\newcommand{\GW}{{\rm GW}}
\newcommand{\LightSeed}{\texttt{Light Seed}}
\newcommand{\HeavySeed}{\texttt{Heavy Seed}}
\newcommand{\UltraLightSeed}{\texttt{Ultra-Light Seed}}

\newcommand{\Msun}{M_{\odot}}

\newcommand{\M}{\mathcal{M}}
\newcommand{\logten}{\log_{10}}
\newcommand{\OmegaAstro}{\Omega_{\GW}^{\rm astro}}
\newcommand{\OmegaCosmo}{\Omega_{\GW}^{\rm cosmo}}
\newcommand{\OmegaGW}{\Omega_{\GW}}
\newcommand{\bn}{\hat{ \mathbf{n}}}

\newcommand{\nn}{\nonumber}

\def\VEV#1{{\left\langle #1 \right \rangle}}
\def\lr#1#2#3{{\left#1 #2 \right#3}}

\def\be{\begin{equation}}
\def\ee{\end{equation}}
\newcommand{\beq}{\begin{eqnarray}}
\newcommand{\eeq}{\end{eqnarray}}

\newcommand{\jhu}{William H. Miller III Department of Physics and Astronomy, Johns Hopkins University, Baltimore, MD 21218, USA}
\newcommand{\ias}{Institute for Advanced Study, 1 Einstein Dr, Princeton, NJ 08540, USA}
\newcommand{\perimeter}{Perimeter Institute for Theoretical Physics, 31 Caroline St N, Waterloo, ON N2L 2Y5, Canada}

\begin{document}

\title{A Fog Over the Cosmological SGWB:\\ Unresolved Massive Black Hole Binaries in the LISA Band}

\date{\today}

\author{Mesut Çalışkan}
\email{caliskan@jhu.edu}
\affiliation{\jhu}

\author{Neha Anil Kumar}
\email{nanilku1@jhu.edu}
\affiliation{\jhu}

\author{Marc Kamionkowski}
\affiliation{\jhu}

\author{Sihao Cheng}
\affiliation{\ias}
\affiliation{\perimeter}


\begin{abstract}
\noindent Disentangling the rich astrophysical structure of the stochastic gravitational-wave background (SGWB) from its cosmological component is essential for the Laser Interferometer Space Antenna (LISA) to access the physics of the early Universe beyond the reach of any other probe.
In this work, we develop an analytical framework to compute the astrophysical contribution to the SGWB arising from an unresolved ensemble of inspiraling and merging black hole binaries.
Accounting for various resolvability thresholds, we leverage this framework to predict the amplitude, spectral shape, and detection signal-to-noise ratio of the unresolved background from massive black hole binaries (MBHBs), capturing the possible diversity of its SGWB imprint across a range of astrophysically motivated populations.
Through a joint analysis of astrophysical and primordial contributions to the SGWB, we determine the minimum detectable amplitude of the cosmological background across a range of spectral shapes.
We demonstrate that, even under optimistic subtraction thresholds, unresolved MBHBs can degrade the detectability of a cosmological signal by multiple orders of magnitude, depending on the spectral shape of the primordial component.
Ultimately, the MBHB-induced astrophysical SGWB acts both as a veil and a lens: it imposes a fundamental limit on cosmological sensitivity, yet simultaneously reveals the hidden population of massive black holes beyond the reach of individual detections. 
Accurate modeling of this late-Universe background is therefore a prerequisite for robust component separation and for realizing LISA's full scientific potential.
\end{abstract}

\maketitle


\section{Introduction}
\label{sec:intro}

Understanding stochastic gravitational-wave backgrounds (SGWBs) and their implications for early-Universe and fundamental physics is a primary science objective of the Laser Interferometer Space Antenna (LISA)~\cite{lisaproposal,LISA:2024hlh}.
Such backgrounds may be of astrophysical (late-time) or cosmological origin\footnote{In this paper, we will refer collectively to all primordial or cosmological SGWBs simply as cosmological backgrounds, without distinguishing between specific subcategories.}, with the latter potentially sourced by inflation~\cite{Grishchuk:1974ny}, axion inflation~\cite{Barnaby:2011qe}, phase transitions~\cite{Kamionkowski:1993fg}, cosmic strings and axionic relics~\cite{Damour:2004kw, Siemens:2006yp, Olmez:2010bi, Regimbau:2011bm}, or pre-inflationary remnants~\cite{Odintsov:2021urx} (see also~\cite{LISACosmologyWorkingGroup:2022jok}).

A precise measurement of the cosmological gravitational-wave (GW) background across multiple frequency bands is essential to constrain inflationary physics and to probe changes in the effective number of relativistic degrees of freedom throughout the Universe's thermal history~\cite{Smith:2005mm,Boyle:2007zx,Smith:2008pf}.
Observational efforts such as measurements of the cosmic microwave background B-mode polarization, pulsar timing arrays (PTAs), and ground-based detectors like Laser Interferometer Gravitational-Wave Observatory (LIGO) target the characterization of this background across widely separated frequency bands. 
LISA will bridge the gap between existing probes by detecting GWs in a previously unexplored frequency regime (roughly from $10^{-5}\, \mathrm{Hz}$ to $0.5\, \mathrm{Hz}$), offering unique insights into epochs inaccessible to electromagnetic observations.

A prominent obstacle to unambiguously detecting the cosmological background is disentangling it from low-redshift, astrophysical contributions to the SGWB.
Previous work has typically considered astrophysical backgrounds from sources like galactic white dwarfs, stellar-mass black hole binaries, neutron star binaries, or extreme mass ratio inspirals (EMRIs)~\cite{Nelemans:2001hp,Sesana:2016ljz,Bonetti:2020jku,Smith:2020lkj,Babak:2023lro, Biscoveanu:2020gds}.
However, in the context of the LISA band, prior studies have largely overlooked the contribution to the SGWB from unresolved massive black hole binary (MBHB) mergers.\footnote{In this work, MBHBs include systems ranging from intermediate-mass black holes ($\sim 250\, \Msun$) to supermassive black hole binaries ($\sim 10^9\, \Msun$).} 

While a substantial fraction of MBHBs in the Universe will produce individually resolvable signals for LISA~\cite{Sesana:2004sp}, some will likely remain unresolved.
For instance, light-seed formation scenarios predict hundreds of MBHB mergers per year, many of which are expected to fall below LISA's detectability threshold~\cite{Barausse:2012fy,Klein:2015hvg,Volonteri:2016uhr,Barausse:2023yrx}.
These unresolved events contribute to the astrophysical SGWB, generated by the cumulative GW emission from an ensemble of individually undetectable, independently evolving systems.
If the ensemble is dominated by binaries in their early inspiral phase, their superposed emission will follow an adiabatically growing power-law signal.
However, since binaries merge within or near the LISA band, their signals may significantly deviate from this regime, giving rise to a more intermittent, ``popcorn''-like background structure.\footnote{We refer to this unresolved component as the astrophysical \emph{background}, though for cosmological signal detection, it can be considered a foreground. Since our aim is to study multiple SGWB contributions jointly, we refer to all components as backgrounds.}

Separating the SGWB induced by late-time sources from potential early-Universe signatures is feasible, provided that a reliable population model for sources contributing to the astrophysical SGWB is available.
However,  despite the extraordinary data obtained from major galaxy surveys and space observatories such as the Hubble Space Telescope~\cite{Lallo:2012xs} and the James Webb Space Telescope~\cite{Gardner:2006ky}, the demographics of high-mass binary systems remain uncertain.
For example, recent detection of a stochastic signal by PTAs in the $\sim\mathrm{nHz}$ band~\cite{NANOGrav:2023gor,EPTA:2023fyk,Reardon:2023gzh,Xu:2023wog} has already sparked debate: some argue it matches predictions from hierarchical models of galaxy formation~\cite{Liepold:2024woa}, while others suggest the signal amplitude exceeds expectations by nearly an order of magnitude~\cite{Sato-Polito:2023spo,Sato-Polito:2023gym}.
Although the MBHB population contributing to the SGWB within the LISA band may be better constrained, significant uncertainties persist, particularly at the lower-mass end~\cite{Caliskan:2022hbu, Caliskan:2023zqm, Langen:2024ygz}, which could dominate the unresolved background.
This view is supported by recent observations of both supermassive black holes~\cite{Melo-Carneiro:2025umm} and novel techniques aimed at identifying numerous (lower-mass) massive black holes~\cite{Grishin:2025nmr}.

In this work, we analyze MBHB population models to quantify the amplitude and spectral shape of the unresolved MBHB-induced SGWB in the LISA frequency band.
We construct a flexible \textit{analytical} framework capable of accurately computing the spectral energy density (SED), accounting for binaries merging within or near LISA’s observational window.
We validate its accuracy through comparison with numerical calculations, employing full inspiral–merger–ringdown (IMR) waveforms.

\textit{Critically, we determine the smallest detectable amplitude of the cosmological SGWB in the presence of an unresolved MBHB background.}
To this end, we present an analytical MBHB population model that offers the flexibility to efficiently reproduce state-of-the-art predictions for merger rates, redshift distributions, and mass functions.
To robustly assess detectability and foreground separation, we perform joint information-matrix analyses across a broad range of cosmological background amplitudes and spectral shapes.
In addition to evaluating the detectability of the astrophysical background itself, we quantify its impact on cosmological signal recovery by demonstrating how the presence of an unresolved MBHB background degrades LISA’s ability to measure cosmological background parameters.
This allows us to delineate the parameter regimes where separation between astrophysical and cosmological backgrounds remains viable.

By considering a broad suite of MBHB formation scenarios spanning different redshift and mass distributions, as well as a wide range of merger rates, our analysis provides a comprehensive characterization of the unresolved SGWB in the LISA band.
This framework not only informs strategies for disentangling cosmological and astrophysical components, but also provides insight into MBHB populations beyond the reach of individual source detection.
In particular, it enables exploration of lighter systems that are too massive to merge in the LIGO band and too faint to be resolved by LISA.
Accurate modeling of the unresolved SGWB is thus essential, not only to improve sensitivity to cosmological signals, but also to probe the formation pathways and evolution of MBHBs, offering a unique perspective on the growth of structure in the Universe.

The paper is organized as follows.
In Sec.~\ref{sec:spectral_density}, we present the formalism used to compute the SED of an SGWB from a population of binary black holes, including both inspiraling and merging sources.
Section~\ref{sec:population_models} introduces our analytical parameterization of the MBHB population and delineates the canonical models used for illustration.
In Sec.~\ref{sec:astrophysical_background}, we present our analytical calculation of the astrophysical background from unresolved MBHBs, describe the algorithm used to determine detectability with LISA, and quantify how model parameters and subtraction thresholds affect the resulting background.
Section~\ref{sec:primordial_background} defines the cosmological background model, specifying the assumed range of spectral indices and amplitudes.
In Sec.~\ref{sec:results}, we present our signal-to-noise ratio (SNR) calculations and information-matrix forecasts for the detectability of both cosmological and astrophysical backgrounds.
We begin by reviewing LISA’s experimental specifications in Sec.~\ref{sec:exp_spec}, followed by our methodology for computing optimal SNRs and constructing the information matrices in Secs.~\ref{sec:how_to_SNR} and~\ref{sec:how_to_Fisher}, respectively.
Section~\ref{sec:SNR_forecasts} presents SNR forecasts for the unresolved MBHB background.
In Sec.~\ref{sec:Fisher_forecasts}, we forecast LISA’s joint sensitivity to a cosmological background in the presence of unresolved MBHBs, determine the smallest detectable cosmological amplitude across various spectral indices, and quantify the associated parameter degeneracies.
We also constrain how the unresolved MBHB background degrades cosmological constraints, and identify the regimes where foreground separation remains viable.
Section~\ref{sec:conclusions} provides concluding remarks.
Throughout, we assume a $\Lambda\textrm{CDM}$ cosmology with cosmological values matching Planck 2018~\cite{Planck:2018vyg}.


\section{GW spectral energy density from a population of binaries}
\label{sec:spectral_density}
The energy density of an astrophysical GW background at detector-frame frequency $f$ is the sum of GW emission from all binaries emitting at that observed frequency. 
Therefore, given a model for the number density of (inspiraling or merging) binary sources, the SED from an ensemble of binaries can be quantified as follows:
\begin{widetext}
\begin{equation}
    \Omega_\GW(f) \equiv \frac{1}{\rho_c}\frac{\dd \rho_\GW}{\dd \ln f} = \frac{1}{\rho_c c^2}\int \frac{\dd z}{(1+z)}\int \dd \logten \M \frac{\dd n}{\dd z\dd \logten\M}\lr{[}{\frac{\dd E_\GW}{\dd \ln f_r}}{]}\Bigg|_{f_r = f(1+z)}\,,
    \label{eqn:fullbackground}
\end{equation}
\end{widetext}
where $f_r$ is the frequency of GWs in the source’s cosmic rest frame, $\Omega_\GW(f)$ is the GW energy density of the ensemble, per logarithmic frequency bin, in units of the critical energy-density of the universe $\rho_c$, and ${\dd \rho_\GW}/{\dd \ln f}$ represents the absolute SED of the population of sources per $\ln f$-bin. 
In the above equation, we have assumed that the comoving number-density of sources $n$ only depends on two parameters: the source-frame chirp mass $\M$ and source redshift $z$. 
The number-density of relevant sources (per $z$ per $\logten \M$) is then translated into the SED expected from each binary system via the energy of GW emission per logarithmic frequency bin, from each individual source,  $\dd E_\GW/\dd \ln f_r$.
It is important to note that, given a realistic prescription for the population of binary sources, and an exact model for $\dd E_\GW/\dd \ln f_r$, which most generally can depend on all source parameters characterizing an inspiraling binary, the above equation exactly quantifies the sky location-, inclination-, polarization angle-, and coalescence phase-averaged signal expected from an ensemble of inspiraling/merging binary systems.
In other words, the only approximation made in the equation above is the simplifying assumption that the population of binary sources can be characterized in terms of only two source parameters $z$ and $\M$.

In this work, we approximate $\Omega_{\rm GW}$ using an analytical model $\dd E_\GW/\dd \ln f_r$. 
When characterizing the background from gradually inspiraling binaries, such as those sourcing GW emission in the PTA band, the function $\dd E_\GW/\dd \ln f_r$ can be described analytically to a good approximation. 
However, the ensemble of MBHBs characterizing GW emission in the LISA band have a much shorter lifetime, evolving more rapidly and often merging within band.
Moreover, the exact $\dd E_\GW/\dd \ln f_r$ per MBHB in the LISA band nominally depends on all the source parameters characterizing the binary system and also requires numerical computation~\cite{Marsat:2020rtl, Lousto:2022hoq}.
In our work, however, the evolution of each MBHB system is approximated using the following equation:
\begin{eqnarray}
    \frac{\dd E_\GW}{\dd \ln f_r} \approx \frac{1}{3G}(G\M)^{5/3}(\pi f_r)^{2/3}\Theta\lr{[}{f_{r, {\rm ISCO}} - f(1+z)}{]}\,,\nonumber\\
    \label{eq:individual_MBHB_SED}
\end{eqnarray}
where $f_{r, {\rm ISCO}}$ is the source-frame frequency corresponding to the inner-most stable circular orbit (ISCO) of the MBHB. 
The above prescription tracks the evolution of an equal-mass (mass ratio $q$ = 1) black-hole binary in a circular orbit until $f_{r, {\rm ISCO}}$, beyond which the GW emission can no longer be computed analytically. 
Instead of attempting to approximate the GW emission beyond the ISCO, when the binary system undergoes a merger, the Heaviside step function $\Theta\lr{[}{f_{r, {\rm ISCO}} - f(1+z)}{]}$ ensures that GW emission from the MBHB is only considered (roughly) until the frequency corresponding to the merger.
In other words, this prescription ignores the dependence of GW emission on any binary source parameters beyond $\M$ and $z$, and more importantly, neglects the GW emission sourced by the merger and ringdown phases of the MBHB. 
Therefore, the $\Omega_{\rm GW}(f)$ computed using the SED model presented in Eq.~\eqref{eq:individual_MBHB_SED} should be considered an approximate, sky-averaged lower-bound on the GW energy density from a given ensemble of binary mergers in the LISA band.\footnote{On the other hand, assuming $q = 1$ maximizes the radiated GW energy at fixed chirp mass; allowing for a distribution of mass ratios with $q > 1$ will slightly reduce the total $\Omega_{\rm GW}(f)$ predicted by the same chirp mass and redshift distribution.}
For our models and forecasts, we adopt $f_{r,{\rm ISCO}} \simeq 4.7 \times 10^{-3} \times (10^6 M_\odot / M)\,{\rm Hz}$, which applies for spinless, equal-mass MBHBs, where $M$ is the total mass of the binary~\cite{Maggiore:2018sht}.
Given the above prescription to approximate $\Omega_{\rm GW}(f)$ and $\dd E_\GW/\dd \ln f_r$, the only remaining ingredient to characterize the expected SED from MBHBs in the LISA band is modeling the expected population of binaries contributing to this signal.


\section{An Analytical MBHB Population Model}
\label{sec:population_models} 
To characterize the population of MBHBs emitting GWs in the LISA band, we begin by adopting a general model for their number density per unit comoving volume, focusing on binaries with chirp masses in the range $250 \,M_\odot \lesssim \mathcal{M} \lesssim 10^9\,M_\odot$ as follows~\cite{Sato-Polito:2023spo, Middleton:2015oda}:
\begin{align}
    \frac{\dd n}{\dd z\,\dd \logten\M} 
    &= \dot{n}_0 
    \left[ \left( \frac{\M}{10^7} \right)^{-\alpha} 
    e^{-\M/\M_*} \right] \notag \\
    &\hspace{4em} \times \left[ (1+z)^\beta e^{-z/z_0} \right] 
    \frac{\dd t_r}{\dd z}\,, 
    \label{eqn:pop_dist}
\end{align}
where $t_r$ is coordinate time in the source's cosmic rest-frame, i.e., $\dd t_r/\dd z = [(1+z)H(z)]^{-1}$, and we have assumed that this distribution is independent of any binary source-parameters besides the source redshift $z$ and chirp mass $\M$. 
Specifically, we assume that all binaries are spinless and equal mass ($q=1$).
For the results in this section, we assume MBHB sources span the redshift range $0.1 \lesssim z \lesssim 20.0$.
Finally, in the above model, we have used five model parameters $\{\alpha,\, \M_*,\, \beta,\, z_*,\, \dot{n}_0\}$, where $\dot{n}_0$ sets the normalization of the merger-rate density. 

Instead of specifying $\dot{n}_0$ directly, it is implicitly determined in terms of the total number of binary mergers expected per year as follows. 
Given the number-density of sources in Eq.~\eqref{eqn:pop_dist}, one can derive the more familiar merger rate per $z$-bin, per logarithmic $\M$-bin, to be
\begin{equation}
     \frac{\mathrm{d}N}{\mathrm{d}z\mathrm{d}\log_{10} \mathcal{M}} = \frac{\mathrm{d}n}{\mathrm{d}z\mathrm{d}\log_{10} \mathcal{M}} \frac{\mathrm{d}V_c}{\mathrm{d}z} \frac{\mathrm{d}z}{\mathrm{d}t}\,,
     \label{eq:merger_rate_definition}
\end{equation}
where $N$ represents the merger rate per unit time $t$ in the detector frame, and $V_c$ represents the comoving volume at redshift $z$. 
The conversion factor, therefore, simplifies to $\dd V_c/\dd z \times \dd z/\dd t = 4\pi\chi(z)^2$, where $\chi(z)$ represents the comoving distance to $z$. 
Then, the total number of mergers per year, across the considered chirp-mass and redshift range of the population, is given by:
\begin{equation}
    N_0 \equiv \int \dd z \int \dd \logten \M  \frac{\mathrm{d}N}{\mathrm{d}z\mathrm{d}\log_{10} \M}\,. 
\end{equation}
This means that, for a fixed $N_0$, the fiducial value of the model parameter $\dot{n}_0$ can be calculated as
\begin{eqnarray}
    \dot{n}_0 = N_0 \lr{[}{\int \dd z \dd \logten \M \frac{\mathrm{d}N}{\mathrm{d}z\mathrm{d}\logten \M}(\M, z| \dot{n}_0 = 1.0)}{]}^{-1}\,,\nonumber\\
    \label{eq:N0_to_n0}
\end{eqnarray}
where the integrand represents the merger rate from Eq.~\eqref{eq:merger_rate_definition} calculated assuming $\dot{n}_0 = 1$.

We emphasize that the simplicity and analytical nature of the population model in Eq.~\eqref{eqn:pop_dist} offers several practical advantages. 
First, the model’s compact parametrization provides the flexibility to emulate a wide range of semi-analytical or simulation-based MBHB population models, while maintaining minimal computational cost.
Second, the separately factorized $\M$ and $z$ dependence in Eq.~\eqref{eqn:pop_dist} enables clear physical interpretation of how different population features, such as the abundance of high-mass binaries or a shift in redshift peak, impact observable quantities like the SGWB amplitude or SNR.
Third, the model is computationally efficient and fully differentiable, making it particularly well suited for parameter estimation, where analytic derivatives with respect to population parameters are desirable.
Finally, the compact form and transparency of the model facilitate reproducibility and cross-comparison with other studies.


The purely analytical framework presented above is used to generate potential MBHB populations by fitting the parameters $\{\alpha,\, \M_*,\,\beta,\,z_0,\, \dot{n}_0\}$ to match predictions from existing literature.
A variety of techniques have been adopted to predict MBHB population statistics, including analytical~\cite{Sesana:2008mz}, semi-analytical~\cite{Barausse:2012fy}, and simulation-based approaches~\cite{Katz:2019qlu, Sesana:2007sh}.
In this work, we choose to calibrate the model parameters to match semi-analytical predictions from Ref.~\cite{Barausse:2023yrx}. These predictions are based on an updated version of the galaxy formation framework developed in Ref.~\cite{Barausse:2012fy}, incorporating recent constraints from PTAs.
This approach retains the computational advantages and interpretability of our analytical model while capturing key astrophysical features of some of the state-of-the-art MBHB population models.

\begin{figure*}[t]
    \centering
    \includegraphics[width=1.0\textwidth]{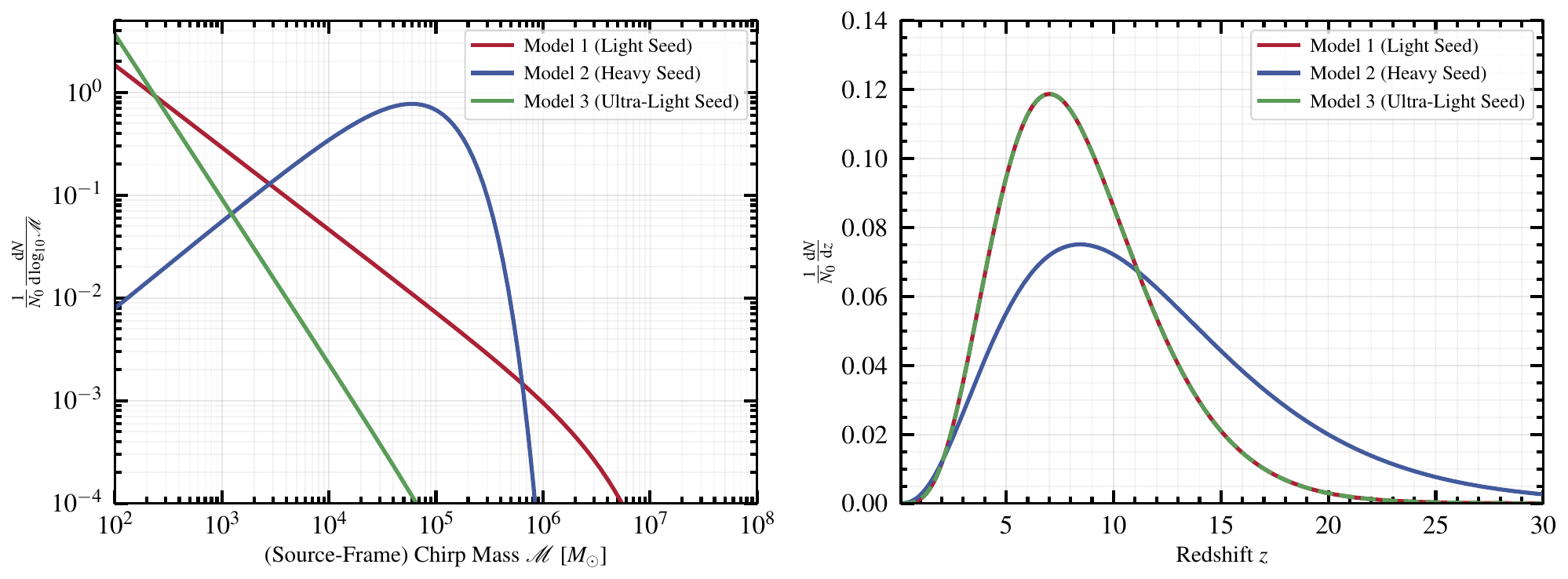} 
    \caption{
    Probability density functions of the logarithmic chirp mass (left panel) and redshift (right panel),
    obtained by integrating the population number density given in Eq.~\eqref{eqn:pop_dist} over all available redshifts (for the mass distribution) or chirp masses (for the redshift distribution). 
    The three curves correspond to \textit{Model 1 (Light Seed)}, \textit{Model 2 (Heavy Seed)}, and \textit{Model 3 (Ultra-Light Seed)} in red, blue, and green, respectively.
    The underlying population parameters for each model are listed in Tab.~\ref{tab:pop_model_params}. 
    Because Models~1 and~3 employ the same $\beta$ and $z_0$, their redshift distributions are identical and thus overlap on the right panel.
    }
    \label{fig:pop_models}
\end{figure*}  

\begin{table*}[t]
    \centering
    \caption{
        Population parameters for the three models used to generate the chirp-mass and redshift distributions in Fig.~\ref{fig:pop_models} and Eq.~(\ref{eqn:pop_dist}). 
        Each model assumes a power-law distribution in chirp mass $\M$ with slope $\alpha$ and characteristic mass scale $\M_*$, 
        and a redshift evolution governed by the parameters $\{\beta,\, z_0\}$. 
        The final column lists the normalization constant $\dot{n}_0$, 
        which sets the overall amplitude of the comoving merger rate density such that the total merger rate $N_0$ integrates to 200~yr$^{-1}$ for each model.
    }
    \label{tab:pop_model_params}
    \vspace{10pt}
    \begin{minipage}{\textwidth}
    \renewcommand{\arraystretch}{1.5}
    \centering
    \small
    \begin{tabular}{l|>{\centering\arraybackslash}p{1.5cm} >{\centering\arraybackslash}p{1.5cm}|
                      >{\centering\arraybackslash}p{1.5cm} >{\centering\arraybackslash}p{1.5cm}|
                      >{\centering\arraybackslash}p{2.5cm}}
        \hline \hline
        Model & \multicolumn{2}{c|}{Mass Parameters} & \multicolumn{2}{c|}{Redshift Parameters} & $\dot{n}_0$ [Mpc$^{-3}$\ yr$^{-1}$] \\ \cline{2-5}
              & $\alpha$ & $\mathcal{M}_*$ $[M_\odot]$ & $\beta$ & $z_0$ & \\ \hline
        Model 1 \texttt{(Light Seed)}       & 0.8     & $5 \times 10^6$  & 7.2 & 1.5 & $7.14 \times 10^{-18}$ \\
        Model 2 \texttt{(Heavy Seed)}       & $-0.85$ & $7 \times 10^4$  & 4.8 & 3.3 & $2.01 \times 10^{-11}$ \\
        Model 3 \texttt{(Ultra-Light Seed)} & 1.6     & $5 \times 10^5$  & 7.2 & 1.5 & $2.97 \times 10^{-21}$ \\
        \hline \hline
    \end{tabular}
    \end{minipage}
\end{table*}

For our forecasts, we consider three MBHB population scenarios.
Model 1 (\texttt{Light Seed}) and Model 2 (\texttt{Heavy Seed}) are anchored to distinct black hole seeding and evolutionary channels identified in Ref.~\cite{Barausse:2023yrx}.
Model~3 (\UltraLightSeed) is synthetically constructed to illustrate that a higher fraction of low-$\M$ binaries does not necessarily result in a stronger background signal.
Figure~\ref{fig:pop_models} visually summarizes the chirp mass and redshift distributions assumed for each of the three population models. 
The left (right) panel shows the source-frame $\M$-dependence ($z$-dependence) of the unit-normalized yearly merger rate for each model.
In each panel, the red, blue and green curves represent the \LightSeed, \HeavySeed, and \UltraLightSeed~parametrization, respectively.

A summary of the fiducial parameter values assigned to $\{\alpha,\, M_*,\, \beta,\, z_0,\, \dot{n}_0\}$, for each of the aforementioned population models, is summarized in Tab.~\ref{tab:pop_model_params}.
The free parameter $\dot{n}_0$ is fixed implicitly via Eq.~\eqref{eq:N0_to_n0} by specifying the number of mergers per year $N_0$.
In the forecasts that follow, each of the MBHB populations is normalized such that $N_0 = 200\,{\rm yr}^{-1}$.
This standardization allows us to isolate the effects of differing $\M$ and $z$ distributions, without conflating them with changes in signal amplitude induced by variations in the total number of mergers. 
Furthermore, we find that many of our forecasts and final results are either weakly dependent on $N_0$ or scale linearly with its assumed fiducial value. 
We therefore fix $N_0$ for clarity and consistency across models.
The astrophysical motivation behind the characteristic features of each model is detailed below, along with the fiducial merger rates one would realistically associate with each. Given that some semi-analytic models predict merger-rates lower than $200\,{\rm yr}^{-1}$, we explore the implications of such low-rate populations in Appendix~\ref{app:flat_spectrum_forecasts} and find that our main results are robust to these changes.

The \LightSeed~scenario assumes that MBHBs form from the remnants of Population III stars in low-metallicity environments at redshifts $15 \lesssim z \lesssim 20$.
This model predicts the formation of binaries that are initially of low mass, with mergers that are delayed by dynamical evolution timescales between host galaxy mergers and final binary coalescence. 
As a result, the red curves in Fig.~\ref{fig:pop_models} depict a broad redshift distribution peaking at $z \sim 7$, and a large number of low-$\M$ MBHBs, with the rate of mergers gradually decreasing as $\M$ increases.
When sampled directly from the semi-analytical model, the total merger rate is approximately $N_0 \sim 180\, \rm{yr}^{-1}$. 

The \HeavySeed~scenario, by contrast, assumes that MBHBs originate from the direct collapse of massive protogalactic disks at redshifts $10 \lesssim z \lesssim 15$, yielding seed masses of $\sim 10^5\,\Msun$ at formation. 
Furthermore, unlike the \LightSeed~scenario, this model does not assume a significant delay between galaxy and binary mergers.
Therefore, the \HeavySeed~model predicts a broader redshift kernel (peaking at $z\sim 8$) and a significantly lower rate of small-$\M$ MBHB mergers.
However, it is important to note that, despite being titled the \HeavySeed~scenario, this model predicts a lower number of mergers at $\M\gtrsim 10^6\Msun$ than the \LightSeed~counterpart.
When sampled from Ref.~\cite{Barausse:2023yrx}, this no-delay, heavy-seed variant fiducially yields $N_0 \sim 120\,{\rm yr}^{-1}$, which we again round to $200\,{\rm yr}^{-1}$ for comparison.\footnote{
The more conservative variant of this model includes the time delays between formation and merger of the MBHBs and results in an order of magnitude lower merger rate.
}

We note that these two formation scenarios are not mutually exclusive.
It is entirely plausible that the LISA MBHB population will include a mixture of binaries from both light-seed-like and heavy-seed-like formation channels. 

Lastly, to demonstrate that populating the low-mass end of the MBHB distribution does not trivially enhance the GW background, we introduce the \texttt{Ultra-Light Seed}~model as a synthetic population. 
This toy model retains the redshift distribution of Model~1 but imposes a steeper chirp-mass slope, thereby populating very light binaries almost exclusively, predicting almost no mergers with $\M \gtrsim 10^5\,\Msun$ over a four-year LISA-like mission.

\begin{figure}[t]
    \centering
    \includegraphics[width=0.5\textwidth]{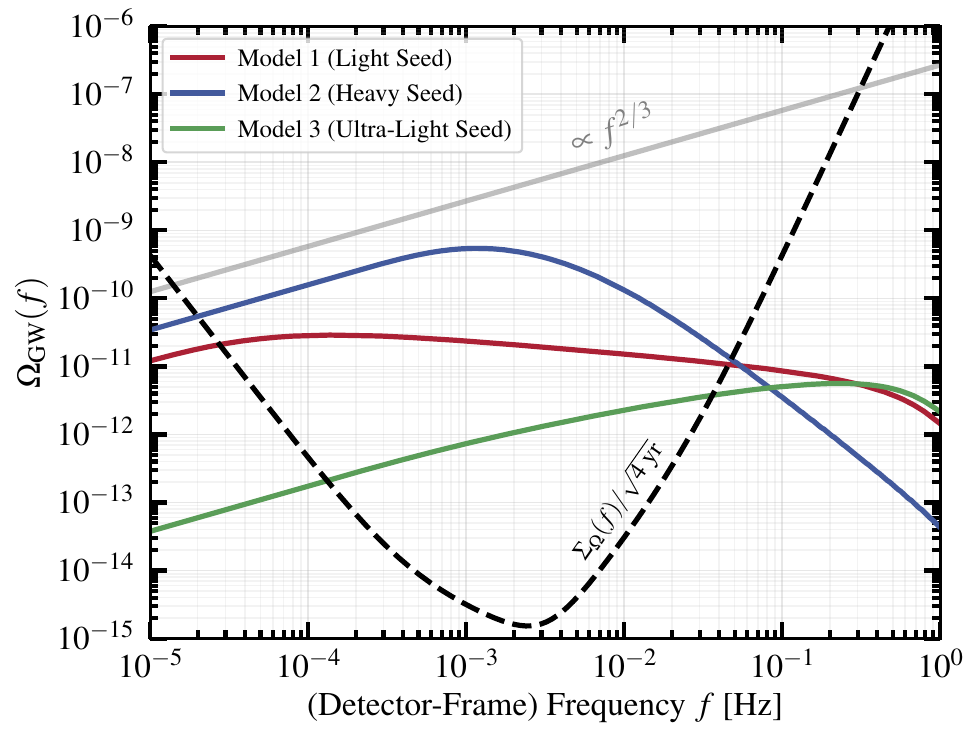} 
    \caption{
    The \textit{total} (resolved + unresolved binaries) spectral GW energy density $\OmegaGW(f)$, calculated using Eq.~\eqref{eqn:fullbackground}, as a function of detector-frame frequency $f$ for the three population models presented in Fig.~\ref{fig:pop_models}. Each population is normalized to yield 200 mergers per year.
    The gray line highlights the characteristic $\propto f^{2/3}$ slope as a benchmark for comparison; mergers occurring in-band cause a deviation from this scaling.
    The black dashed curve represents $\Sigma_{\Omega}(f)$, the nominal sensitivity function of LISA given in Eq.~\eqref{eq:Sigma_Omega_f}, which accounts for LISA's detector response (scaled down by $T_{\rm obs}^{1/2} = \sqrt{4\,{\rm yr}}$).
    The SNRs computed using Eq.~\eqref{eq:SNR_of_OmegaGW} for a 4-year LISA mission are $\sim 666$, $\sim 14473$, and $\sim 43$ for Model~1~(\LightSeed), Model~2~(\HeavySeed), and Model~3~(\UltraLightSeed), respectively.
    }
    \label{fig:Omega_GW_Total_Models}
\end{figure}

Given this prescription for the three population models, we finally use Eq.~\eqref{eqn:fullbackground} to compute the \textit{total} $\OmegaGW(f)$ expected from \textit{all} (resolved + unresolved) MBHB sources for each population model. 
The results are presented in Fig.~\ref{fig:Omega_GW_Total_Models}. 
Consistent with Fig.~\ref{fig:pop_models}, the red, blue, and green curves represent the \LightSeed, \HeavySeed, and \UltraLightSeed~models, respectively.
For comparison, the $\Omega_{\rm GW}(f) \propto f^{2/3}$ spectrum expected from a population of inspiraling (non-merging) sources is shown in gray. 
The LISA sensitivity curve (described in Sec.~\ref{sec:how_to_SNR}), scaled down by the square root of the total observation time $T_{\rm obs}^{1/2} = \sqrt{4\,{\rm yr}}$, is plotted in black to reflect the effective instrument noise.

Figure~\ref{fig:Omega_GW_Total_Models} indicates that the strongest \textit{total} GW spectral energy density is produced by the \HeavySeed~model (Model 2), due to the high number of MBHBs with $10^4\Msun \lesssim \M \lesssim 10^6\Msun$ that emit more GW energy than less massive binaries. 
However, the SED of the \HeavySeed~population drops steeply at high frequencies, as higher mass MBHBs merge and stop contributing to the background.
At low frequencies, the predicted $\Omega_{\rm GW}(f)$ from this model scales as $f^{2/3}$, as expected for a background dominated by inspiraling binaries. 
This behavior arises because the \HeavySeed~model contains few MBHBs with $\M \gtrsim 10^6\, \Msun$, which would otherwise induce deviations from the $f^{2/3}$ slope. 
Instead, the population is dominated by lower-mass binaries ($\M \lesssim 10^6\, \Msun$) that only merge at $f \gtrsim 10^{-3}$ Hz, leading to a suppression of $\Omega_{\rm GW}(f)$ at higher frequencies.

The \LightSeed~ scenario predicts a lower $\Omega_{\rm GW}(f)$ at frequencies $f \lesssim 10^{-1}$ Hz.
This suppression arises from the significantly smaller number of MBHBs with $10^4\,\Msun \lesssim \M \lesssim 10^6\,\Msun$ produced in this model.
In this low-frequency regime, this model also presents an inflection at $10^{-5}$ Hz.
This feature can be attributed to MBHBs with $\M \gtrsim 10^7\Msun$ that merge at low $f$, causing the first deviation from the $f^{2/3}$ spectrum expected from inspiraling sources. 
In contrast, at $f \gtrsim 10^{-1}$ Hz, where MBHBs with $\M \gtrsim 10^3\,\Msun$ have already merged, the \LightSeed~ SED exceeds that of its \HeavySeed~ counterpart.
This is sustained by the high abundance of low-mass MBHBs in the \LightSeed~ population, which continues to support the amplitude of $\Omega_{\rm GW}(f)$ toward higher frequencies.

Finally, the \UltraLightSeed~scenario predicts the weakest \textit{total} $\Omega_{\rm GW}(f)$ at low frequencies, with signal strength comparable to the \LightSeed~ model at high frequencies ($f\gtrsim 10^{-1}$ Hz).
Because this population model is primarily comprised of MBHBs with $\M\lesssim 10^5 \Msun$, the total predicted SED from this model follows the $f^{2/3}$ for $f\lesssim 10^{-1}$ Hz. 
At $f\sim 0.5$ Hz, where the all the considered low-$\M$ MBHBs cross their $f_{r,{\rm ISCO}}$, the lower bound on chirp mass for all populations causes a steeper cutoff. 

In the above description of the \textit{total} $\OmegaGW(f)$ expected from each population model, we focus our attention on the impact of the chirp-mass distribution attributed to each model, without considering the redshift variation in detail. 
A more thorough analysis of the dependence of $\OmegaGW(f)$ on redshift distribution parameters $\beta$ and $z_0$ is presented in Appendix~\ref{sec:varying_redshift_params}, where we demonstrate that changing the redshift distribution has only a mild effect on the resulting GW background.


\section{Calculating the Astrophysical SGWB}
\label{sec:astrophysical_background}

Although the above formalism (approximately) captures the sky-averaged GW energy density expected from all MBHB sources, the resulting $\Omega_{\rm GW}(f)$ does not represent the astrophysical `background' in the LISA band. 
This is because Eq.~\eqref{eqn:fullbackground} gives the GW energy density from \textit{all} mergers, independent of their detectability. 
However, in reality, for a given MBHB source population, only a certain fraction of the (merging or inspiraling) sources are expected to be individually resolvable. 
MBHBs merging at low redshifts, or with source-frame total masses in the range $10^{4}\, \Msun \lesssim M \lesssim 10^{7}\, \Msun$, will likely be individually resolved (up to $z \lesssim 20$) and removed from the total strain observed by a LISA-like detector. 
As a result, the effective astrophysical background in the LISA band $\OmegaAstro(f)$ consists of the population of unresolved MBHBs that are difficult to individually subtract from the total observed signal.

\begin{figure}[t]
    \centering
    \includegraphics[width=0.5\textwidth]{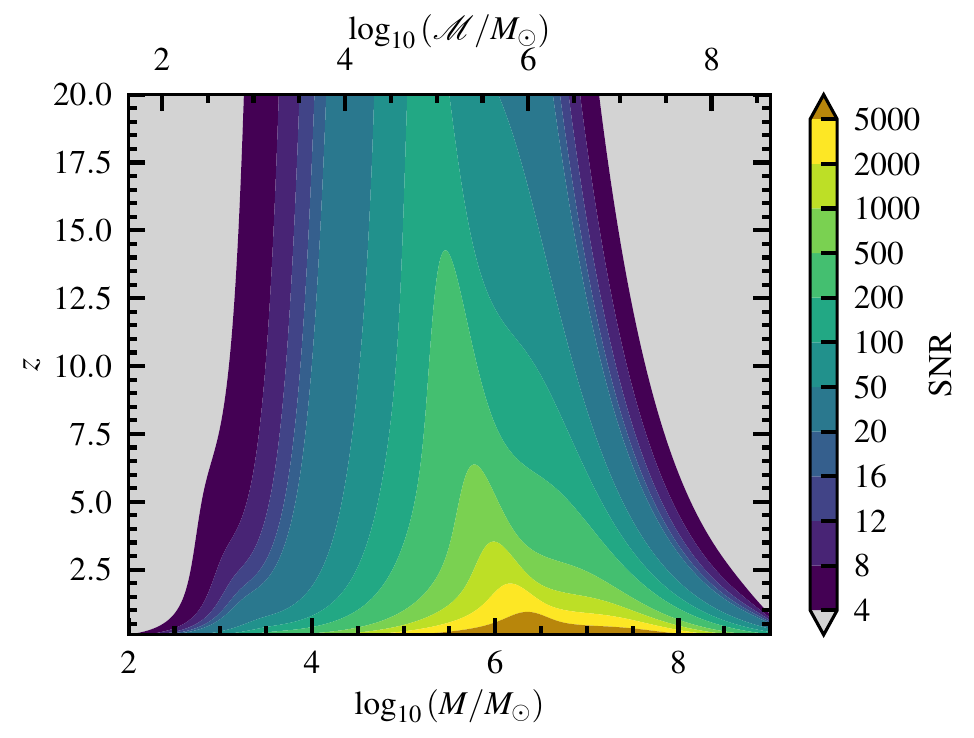} 
    \caption{
    Sky-, inclination-, polarization-, and coalescence phase-averaged SNR for non-spinning ($\chi_1 = \chi_2 = 0)$, equal-mass $(q=1)$ binaries as a function of logarithmic (source-frame) total mass $M$ and redshift $z$.
    The top horizontal axis shows the corresponding logarithmic (source-frame) chirp mass $\mathcal{M}$.
    The contour levels represent SNR boundaries explicitly indicated by the colorbar, with any values below 4 and above 5000 highlighted distinctly in gray and gold, respectively.
    }
    \label{fig:SNR_contour}
\end{figure}

\subsection{Subtracting the resolved binaries}
To obtain the spectral density sourced specifically by unresolved mergers---binaries with an SNR falling below a specified detection threshold---we must first `subtract' the contributions from MBHB systems that are expected to be individually resolvable by a LISA-like detector.
To do so, we start by estimating the optimal SNRs expected from individual MBHBs with a variety of source-frame chirp masses $\M$ and source redshifts $z$, consistent with the bounds on the previously discussed populations.
To calculate the SNRs, we use the \texttt{IMRPhenomX} waveform model~\cite{Pratten:2020fqn} and utilize the \texttt{lisabeta} code~\cite{Marsat:2018oam}, which accounts for the full detector response of LISA.
Although mass and distance are the primary parameters determining the SNR of a binary, other parameters such as spins, sky location, inclination angle, polarization angle, time of coalescence, and coalescence phase can also affect the resulting SNR. 
To account for these variations, the computed SNRs are averaged over the MBHB sky-location, inclination, polarization and coalescence phase.
For simplicity, and due to the uncertainty in the spin distributions, we assume equal-mass binaries and set spins to zero. 
It is important to note we calculate the SNR from each source by injecting it individually into the simulated, LISA-observed signal. 
In other words, the computed SNRs do not account for any `confusion' noise induced by the overlap of this MBHB signal with other binaries inspiraling or merging within the LISA band, including the well-known white-dwarf binary background signal expected to contaminate the observed data set.

Figure~\ref{fig:SNR_contour} displays the sky-, inclination-, polarization-, and coalescence-phase-averaged (individual binary) SNR contours for equal-mass, non-spinning binaries, plotted over a dense two-dimensional grid in redshift $z$ and source-frame total mass $M$.
The top axis indicates the corresponding (source-frame) chirp mass $\M$, facilitating direct comparison with our population model. 
Regions of parameter space with $\mathrm{SNR} < 4$ and $\mathrm{SNR} > 5000$ are explicitly masked in gray and gold, respectively.
By interpolating the contours in Fig.~\ref{fig:SNR_contour}, we can estimate the maximum redshift out to which an MBHB with source-frame total mass $M$ can be individually resolved with a network SNR above a chosen threshold $\rho_{\rm th}$.

Given the SNR interpolation methodology described above, the astrophysical background $\OmegaAstro(f)$ sourced by \textit{unresolved} compact binaries in the LISA band is finally calculated for a given population model as follows:

\begin{widetext}
\begin{equation}
    \Omega_\GW^{\rm astro}(f) = \frac{1}{\rho_c c^2} \int_{\log_{10}\M_{\rm min}}^{\log_{10}\M_{\rm max}}\dd \logten \M \int_{z_{\rm min}(\M,\,\rho_{\rm th})}^{z_{\rm max}} \frac{\dd z}{(1+z)} \frac{\dd n}{\dd z\dd \logten\M}\lr{[}{\frac{\dd E_\GW}{\dd \ln f_r}}{]}\Bigg|_{f_r = f(1+z)}\,,
    \label{eqn:final_astro_background}
\end{equation}
\end{widetext}
where the lower limit of the redshift integral $z_{\rm min}(\M, \rho_{\rm th})$ is determined from the interpolated contour as the maximum redshift out to which an MBHB with source-frame total mass $M = 2^{6/5}\M$ can be individually detected with $\textrm{SNR} \geq \rho_{\rm th}$. 
The remaining integration limits remain unchanged by the application of the SNR cutoff, i.e., we continue to assume $\M_{\rm min} = 250\, \Msun$, $\M_{\rm max} = 10^9\Msun$, and $z_{\rm max} = 20$.

\begin{figure*}[t]
    \centering
    \includegraphics[width=1.0\textwidth]{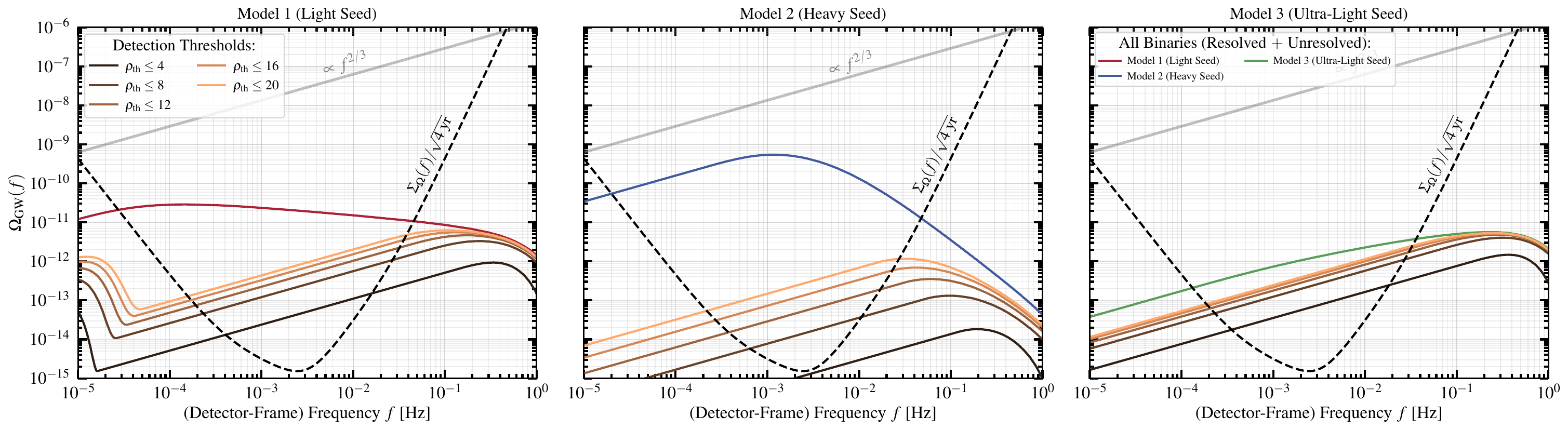} 
    \caption{
    Differential energy density spectra $\Omega_{\mathrm{GW}}(f)$ for three MBHB population scenarios, shown in the left, middle, and right panels for Models~1 (\LightSeed), 2 (\HeavySeed), and 3 (\UltraLightSeed), respectively (as defined in Tab.~\ref{tab:pop_model_params}). 
    The red, blue, and green curves represent the \textit{total} gravitational-wave signal, which includes contributions from both resolved and unresolved binaries (i.e., without applying any detection threshold, $\rho_{\mathrm{th}}<\infty$). 
    Curves in progressively lighter tones depict the stochastic background from unresolved sources, obtained by excluding resolvable mergers above SNR thresholds of $\rho_{\mathrm{th}} = 4,\, 8,\, 12,\, 16,$ and $20$, using a consistent color map across all models.
    Although Model~2 yields the highest total SNR over a 4-year LISA mission, its unresolved background is weaker than that of Model~1, as most binaries in Model~2 are individually detectable. 
    Conversely, Model~3 consists predominantly of unresolved low-mass binaries; however, due to their lower intrinsic GW power, the resulting background is weaker than that of Model~1 (for most detection thresholds) and shows relatively weaker variation with respect to the detection threshold.
    The corresponding SNRs of the residual stochastic backgrounds for each model and threshold are reported in Tab.~\ref{tab:snr_background_models}. 
    The dashed black curve shows the nominal LISA sensitivity $\Sigma_{\Omega}(f)$ (scaled by $T_{\rm obs}^{-1/2}$ to account for a 4 year observation run), and the gray line indicates an $f^{2/3}$ reference slope for comparison.
    }
    \label{fig:Omega_GW_background_allModels}
\end{figure*}

\subsection{The resulting astrophysical SGWBs}

For each of the population models detailed in Sec.~\ref{sec:population_models}, we use the model-agnostic SNR contours presented in Fig.~\ref{fig:SNR_contour} to calculate $\OmegaAstro(f)$. 
Given the uncertainty surrounding the resolvability of individual MBHB signals in the LISA-observed data set using the `Global-Fit' method~\cite{Katz:2024oqg}, we present predictions for $\OmegaAstro(f)$ under various values of $\rho_{\rm th}$.

The three panels in Fig.~\ref{fig:Omega_GW_background_allModels} display our analytical estimates for the sky-averaged $\OmegaAstro(f)$ expected from the \LightSeed, \HeavySeed, and \UltraLightSeed~models, shown from left to right, respectively.
Each panel shows a set of five orange-toned gradient curves representing $\OmegaAstro(f)$ computed using Eq.~\eqref{eqn:final_astro_background} for $\rho_{\rm th} \in \{4,\,8,\,12,\,16,\,20\}$.
The lightest curve corresponds to the strictest individual detection threshold, and the darkest to the most lenient.
For comparison, each panel also displays the previously computed \textit{total} (resolved + unresolved) $\Omega_{\GW}(f)$, including the sky-averaged signal from \textit{all} MBHBs in the corresponding population. 
Finally, each grid also displays an $\Omega_{\GW} \propto f^{2/3}$ spectrum (gray) for comparison, as well as the LISA sensitivity curve (dashed-black), scaled down by $T_{\rm obs}^{1/2} = \sqrt{4\,{\rm yr}}$, for reference.
Details on the sky-averaged LISA sensitivity curve can be found in Sec.~\ref{sec:exp_spec} and~\ref{sec:how_to_SNR}.

\textbf{\textit{On the overall magnitude of the astrophysical background.}} A cursory comparison of the predicted $\OmegaAstro(f)$ spectrum, across all the displayed SNR thresholds, reveals that the \LightSeed~and \UltraLightSeed~population models predict a stronger signal than their \HeavySeed~counterpart. 
Although the \HeavySeed~scenario predicts a significantly stronger \textit{total} $\Omega_\GW(f)$, owing to the its large high-$\M$ MBHB merger content, the inversion in $\OmegaAstro(f)$ signal strength arises because binaries with $10^4\Msun \lesssim \M \lesssim 10^6\Msun$ are individually resolvable by a LISA-like detector out to very high redshifts. 
In other words, since the \LightSeed~and \UltraLightSeed~models contain a much larger fraction of low-$\M$ binary mergers---which are more difficult to resolve individually at $z \gtrsim 2$ (see Fig.~\ref{fig:SNR_contour})---the unresolved background they predict is correspondingly stronger. 

\textbf{\textit{On the deviation from the $\propto f^{2/3}$ spectrum.}} Since the astrophysical background in the LISA band arises from binaries that merge (roughly) within the observable frequency range, it is important to consider how much this signal might deviate from the characteristic $\propto f^{2/3}$ spectrum expected from an ensemble of purely inspiraling binaries.
Contrary to the more pronounced deviation(s) visible in the \textit{total} $\Omega_\GW(f)$ curves for the \LightSeed~and \HeavySeed~scenario, all three models predict a background $\OmegaAstro(f)$ signal that scales mostly as $f^{2/3}$ for a large fraction of the displayed frequency band. 

The \UltraLightSeed~scenario experiences the most modest change in the scaling of the predicted SED with the application of the SNR cutoff. 
This is attributed to the fact that this population is dominated by low $\M$ MBHBs that merge at high frequencies and remain largely unresolved across all the considered values of $\rho_{\rm th}$. 
Thus, the deviation from the $\propto f^{2/3}$ occurs in the same regime ($10^{-1}\ \textrm{Hz} \lesssim f \lesssim 1$\ \textrm{Hz}) for both the \textit{total} $\Omega_\GW(f)$ as well as the background $\OmegaAstro(f)$ signal. 

In contrast, the \HeavySeed~scenario transforms from a spectrum that very visibly deviates from $\propto f^{2/3}$ at frequencies corresponding to the trough of the LISA sensitivity curve, to one that deviates only closer to $10^{-1}\ \textrm{Hz}$ upon the subtraction of the resolved binaries. 
This can, once again, be attributed to the large fraction of individually resolvable MBHBs ($10^4\Msun \lesssim \M \lesssim 10^6\Msun$) that characterize this population model.
The subtraction of these mergers from the total signal leads to a steep drop in overall amplitude and shifts the spectral deviation to higher frequencies, where the remaining low-$\M$ binaries are expected to merge.

Finally, the \LightSeed~model interestingly presents the most significant deviation from the $\propto f^{2/3}$ spectrum. 
On one hand, this population model also predicts an $\propto f^{2/3}$ spectrum for $10^{-4}$ Hz$\lesssim f \lesssim 10^{-1}$ Hz, with a deviation above these frequencies induced by the initial inspiral and following merger of MBHBs with $\M \lesssim 10^4\Msun$.
However, the $\OmegaAstro(f)$ computed for the \LightSeed~scenario also features an additional increase in signal at $10^{-5}$ Hz$\lesssim f \lesssim 10^{-4}$ Hz, causing a deviation from the $\propto f^{2/3}$ spectrum that does not occur in the other considered population scenarios. 
This feature is attributed to the high-mass MBHBs ($10^7\Msun \lesssim \M$) that merge at very low frequencies, and produce a significant amount of GW energy with per source [see Eq.~\eqref{eq:individual_MBHB_SED}].  
The reason this feature is only visible in the \LightSeed~scenario is because this population model predicts a \textit{marginally} higher number-density of very massive mergers ($\M \gtrsim 10^7\Msun$) that are less likely to occur in the \HeavySeed~or \UltraLightSeed~scenarios. 
Even though, in absolute terms, the probability of these high-$\M$ MBHB mergers occurring in the \LightSeed~scenario is very low, this suppression in number-density is outweighed by the magnitude of energy emitted per event. 
It is, however, important to note that `resolvability' in our analytical model is closely tied to the SNR contours presented in Fig.~\ref{fig:SNR_contour}.
But these contours do not account for measurements of higher-order-modes.
Access to modes beyond the quadrupole moment of the GW signal may improve resolvability of individual events, particularly for high-$\M$ systems that merge in the $10^{-5}$ Hz $\lesssim f\lesssim 10^{-4}$ Hz regime. 
In other words, the inclusion of higher-order-modes in the SNR contour computation may erase the existence of this feature even in the \LightSeed~scenario.

\textbf{\textit{On the effect of changing detection threshold.}}
Given the uncertainty in the minimum SNR required by a LISA-like detector to resolve individual MBHB sources, it is instructive to consider how the chosen $\rho_{\rm th}$ affects the amplitude of the predicted $\OmegaAstro(f)$.
For the \LightSeed~and \UltraLightSeed~populations, changing $\rho_{\rm th}$ has a more modest impact on the strength of $\OmegaAstro(f)$.
This can be understood by examining the chirp mass distributions in Fig.~\ref{fig:pop_models}.
As the threshold increases, resolvability becomes increasingly restricted to binaries in the range $10^4\,\Msun \lesssim \M \lesssim 10^7\,\Msun$, pushing more systems into the unresolved regime. 
However, because the \LightSeed~and \UltraLightSeed~populations are dominated by low-$\M$ binaries, whose abundance falls with increasing $\M$, this shift adds only marginally to the unresolved background.
In contrast, the \HeavySeed~scenario is more sensitive to the choice of $\rho_{\rm th}$, as its MBHB population includes a growing number of systems in the range $10^2\,\Msun \lesssim \M \lesssim 10^5\,\Msun$.
Here, increasing $\rho_{\rm th}$ significantly alters the fraction of binaries that are unresolved, leading to a more pronounced change in $\OmegaAstro(f)$.

In summary, a residual astrophysical SGWB persists in the LISA band across all considered population models and most detection thresholds.
This signal is especially robust in the \LightSeed~scenario, indicating that an unresolved background from MBHB mergers is likely to persist in the data, regardless of the efficiency of subtraction techniques.


\section{Cosmological SGWB}
\label{sec:primordial_background}

In the $\sim\!\mathrm{mHz}$ band, where LISA is most sensitive, the leading early-Universe mechanisms predict spectra that can be approximated locally by a power law: nearly scale-invariant ($\gamma \simeq 0$) for inflationary vacuum fluctuations~\cite{Starobinsky:1979ty, Guzzetti:2016mkm}, mildly blue ($\gamma \sim 1$) for networks of cosmic strings~\cite{Auclair:2019wcv, Blanco-Pillado:2024aca}, and steeply rising (falling) with $\gamma \sim 3$ ($-3$) on the low- (high-)frequency side for first-order phase transitions~\cite{Caprini:2019egz}.

Guided by these benchmarks, we model the cosmological background as  
\begin{equation}
\label{eq:Omega_cosmo_PL}
\Omega_{\mathrm{GW}}^{\mathrm{cosmo}}(f)=A
\left(\frac{f}{f_0}\right)^{\gamma},
\qquad f_0 = 2\ \mathrm{mHz},
\end{equation}
where $A\equiv \Omega_{\mathrm{GW}}^{\mathrm{cosmo}}(f_0)$ is the amplitude at the pivot frequency $f_0$, and $\gamma$ is the spectral index.\footnote{The choice $f_0=2\ \mathrm{mHz}$ roughly coincides with the strain-sensitivity minimum of the current LISA design.}

Although real primordial spectra can be more complex, this two-parameter template is flexible enough to encompass many theoretical scenarios while remaining analytically tractable, enabling large-scale parameter scans at negligible computational cost. 
For the forecasts and discussions that follow, we use the terms \emph{primordial} and \emph{cosmological} SGWB interchangeably to refer to the GW background induced by early-Universe physics, reserving the label \emph{astrophysical} for the SGWB generated by unresolved MBHBs in the LISA band.

Throughout our analysis, we treat $(A, \gamma)$ as free parameters and explore
\begin{equation*}
\gamma \in [-5,5], \qquad A \in [10^{-16}, 10^{-10}],
\end{equation*}
which comfortably encloses the projected discovery space of a four-year LISA mission~\cite{Caprini:2015zlo,LISA:2022yao}.
As we will see in the coming sections, the absolute (as opposed to fractional) uncertainty on $A$ depends only on instrument noise, observation time, and the level of astrophysical backgrounds (which act as confusion noise), so surveying such a wide range of amplitudes is very efficient and exhaustive.


\section{Forecasts}
\label{sec:results}

In this section, we present our SNR calculations and information-matrix forecasts for the measurability of both cosmological and astrophysical stochastic backgrounds.  
Section~\ref{sec:exp_spec} reviews LISA’s experimental specifications.  
Section~\ref{sec:how_to_SNR} details our methodology for computing the optimal SNR of a given SGWB, while Sec.~\ref{sec:how_to_Fisher} describes the construction of the information matrix used to forecast uncertainties on the cosmological amplitude $A$, spectral index $\gamma$, and the astrophysical merger rate $N_0$.
A key aim of the information-matrix analysis is to determine the minimum cosmological background amplitude that can be detected in the presence of confusion noise from the unresolved MBHB population. 
That is, we quantify the smallest $A$ that remains distinguishable for each value of $\gamma$, given the computed astrophysical backgrounds.
Our SNR and information-matrix results are presented in Secs.~\ref{sec:SNR_forecasts} and~\ref{sec:Fisher_forecasts}, respectively.  
Finally, in Sec.~\ref{sec:the_numerical_check_forecasts}, we validate our analytic framework by numerically generating $\Omega^{\rm astro}_{\mathrm{GW}}(f)$ from Monte-Carlo realizations of MBHB populations, computing their full waveforms, and comparing the resulting SGWB spectra and SNR forecasts with the analytic predictions.

\subsection{Experiment Specifications}
\label{sec:exp_spec}

Given the simplified model assumptions for the three MBHB populations in our forecasts, and the uncertainty in expected merger rates within the LISA band, we adopt the noise and response specifications from Ref.~\cite{Smith:2019wny} to characterize the sensitivity of a LISA-like detector to the SGWB. 
The noise and response functions are summarized below, along with the detector specifications relevant to our modeling assumptions. 
For a detailed derivation of these expressions, we refer the reader to Ref.\cite{Smith:2019wny}.

The response function to the intensity of a stationary SGWB for a LISA-like detector, most  generally, is a time-dependent quantity due to the orbital motion of the spacecraft. 
However, in our forecasts, we adopt a simplified model that neglects this time dependence, as its impact on the SNR and parameter estimation is expected to be negligible in comparison to the larger uncertainty surrounding the astrophysical population of MBHBs.
Specifically, we assume that the spacecraft positions are fixed in space, with the following coordinates:
\begin{equation}
\begin{aligned}
    \vec{x}_1 &= \{0,\,0,\,0\}\,, \\
    \vec{x}_2 &= L\left\{\tfrac{1}{2},\, \tfrac{\sqrt{3}}{2},\, 0\right\}\,, \\
    \vec{x}_3 &= L\left\{-\tfrac{1}{2},\, \tfrac{\sqrt{3}}{2},\, 0\right\}\,,
\end{aligned}
\end{equation}
where the subscripts $ 1,\, 2 $, and $ 3 $ label the three detector vertices, and $ L = 2.5 \times 10^9\,\mathrm{m} $ is the arm length of the triangular configuration.
Under these assumptions, Ref.~\cite{Smith:2019wny} provides the following approximate fit for the LISA detector response to the SGWB intensity:
\begin{eqnarray}
    \mathcal{R}_{A} &= &\mathcal{R}_E \simeq \frac{9}{20}|W|^2\left[1 + \left(\frac{3f}{4f_*}\right)^2\right]^{-1}\,, \label{eq:A_E_response_functions}\\ 
    \mathcal{R}_T &\simeq &\frac{1}{4032}\left(\frac{f}{f_*}\right)^6 |W|^2\left[1 + \frac{5}{16128}\left(\frac{f}{f_*}\right)^8\right]\,,  
\end{eqnarray}
 where $A,\, E$ and $T$ are none other than the time-delay interferometry (TDI) variables, and the characteristic frequency $f_* \equiv c/(2\pi L)$. Setting $W = 1$ gives the response of the detector to the Michelson signal, whereas, using $W(f,f_*) = 1 - e^{-2if/f*}$ returns the more optimal TDI signal that relies on longer paths to detect phase shifts. 
 Although we maintain this $W$-dependence for generality, in the idealized scenario considered for our forecast calculations, the factors of $W$ for the extra TDI paths cancel. 
 That is, the forecasts presented in this work are not dependent on the functional form attributed to $W$.

The noise model presented in Ref.~\cite{Smith:2019wny} is simplified under the assumption that acceleration noise and optical path-length fluctuations are the most dominant effects, with rms amplitudes
\begin{align}
    \sqrt{(\delta a)^2} = 3\times 10^{-15}\,{\rm{m/s}}^2, && \sqrt{(\delta x)^2} = 1.5 \times 10^{-11}\,{\rm m}.
\end{align}
The acceleration and optical-metrology noise spectra are then given by
\begin{eqnarray}
    S_a &= &\left(\sqrt{(\delta a)^2} / L\right)^2 \frac{[1 + (f_1/f)^2]}{(2\pi f)^4}\,{\rm Hz}^{-1}\,, \\
    S_s &= &\left(\sqrt{(\delta a)^2} / L\right)^2\,{\rm Hz}^{-1}\,,
\end{eqnarray}
where $f_1 = 4\times 10^{-4}$ Hz. 
The noise in the $A$ and $E$ channels can then be approximated as
\begin{eqnarray}
    N_A = N_E \simeq (6S_s + 24S_a)|W|^2\,,
    \label{eq:A_E_noise_spectra}
\end{eqnarray}
where we assumed that the noise power spectrum remains the same across the two TDI channels and the approximation in the above equation is valid in the low frequency regime [$\cos(f/f_*) \simeq 1$]. 
This approximation maintains a good fit to the exact noise curve, without the high-frequency wiggles. 
Because the $T$ mode (also called the Sagnac signal) is characterized by a response function that is minimally sensitive to the SGWB at low frequencies, in comparison to its $A$ and $E$ counterparts, for our forecasts we assume that this channel can be used to entirely characterize the noise spectra associated with the $A$ and $E$ modes above. 
In other words, we solely rely on the $A$ and $E$ mode measurements to characterize the SGWB signal, and use the $T$ channel only to reconstruct the instrument noise spectra $N_A(f)$ and $N_E(f)$. Finally, for all the forecasts presented in this paper, we assume that measurements are made spanning the frequency range $f_{\rm min} = 10^{-5}\ \textrm{Hz}$ to $f_{\rm max} = 0.5\ \textrm{Hz}$. All our forecasts assume a nominal four-year mission, i.e., $T_{\rm obs} = 4$~yr. 

\subsection{How to Calculate the SGWB SNR}
\label{sec:how_to_SNR}
The SNR characterizing the sensitivity of a LISA-like detector to a stationary SGWB, subject to the simplifying assumptions detailed in Sec.~\ref{sec:exp_spec}, is given by the well-known expression~\cite{Maggiore:2007ulw}
\begin{eqnarray}
    {\rm SNR} = \left[T_{\rm obs} \sum_{i\in\{A,\, E\}}\int_{f_{\rm min}}^{f_{\rm max}}\dd f \frac{\mathcal{S}_i^2(f)}{N_i^2(f)}\right]^{1/2}\,,
\end{eqnarray}
where $\mathcal{S}_i \equiv \mathcal{R}_iS_h(f)$ is the \textit{observed} signal power, accounting LISA's response function to the \textit{one-sided} SGWB spectral density
\begin{eqnarray}
    \VEV{h^*_p(f, \bn)h_{p'}(f', \bn')} = \delta(f-f')\frac{\delta^2(\bn, \bn')}{4\pi}\delta_{pp'}\frac{1}{2}S_h(f)\,.\nn\\
    \label{eq:PS_definition}
\end{eqnarray} 
In the above equation, the $\delta$'s represent the Dirac delta function, index $p\in\{+,\, \times\}$ labels the polarization, and $h_+(f, \bn)$ and $h_\times(f, \bn)$ therefore represent the frequency-domain SGWB strain at sky-location $\bn$. 
In writing the above expression, we have continued to assume that the background is stationary and isotropic. 
At this stage, we also restrict attention to the two-point statistics of the isotropic background, ignoring any non-Gaussian (spatial or temporal) features and forecasting sensitivity solely to the variance of the SGWB. 
Note that the factor of $1/(4\pi)$ is a conventional normalization choice, which disappears upon integrating over all directions $\hat{\mathbf{n}}$ and $\hat{\mathbf{n}}'$.

Next, to express this SNR in terms of $\Omega_{\rm GW}(f)$, we use the relation:
\begin{equation}
    \Omega_{\GW}(f) = \frac{4\pi^2f^3}{3H_0^2} S_h(f)\,,
    \label{eq:Omega_GW_to_Sh}
\end{equation}
which applies regardless of the background type, whether $\Omega_{\GW}(f)$ is astrophysical ($\OmegaAstro(f)$), cosmological ($\OmegaCosmo(f)$), or a combination of both. 
This relation then allows us to re-write the inverse noise-weighted response to the variance of the SGWB spectral density as 
\begin{eqnarray}
\label{eq:background_SNR}
    {\rm SNR} = \left[T_{\rm obs} \int_{f_{\rm min}}^{f_{\rm max}}\dd f \frac{\Omega_{\GW}^2(f)}{\Sigma_{\Omega}^2(f)}\right]^{1/2}\,,
    \label{eq:SNR_of_OmegaGW}
\end{eqnarray}
where the effective noise spectrum in the above equation is defined as:
\begin{eqnarray}
    \Sigma_{\Omega}(f) &= &\frac{4\pi^2}{3\sqrt{2}} \frac{f^3}{H_0^2}\frac{N_A}{\mathcal{R}_A}\,, \\
    &\simeq &\frac{4\pi^2f^3}{3H_0^2}\frac{20\sqrt{2}}{3}\left[4S_a(f) + S_s(f)\right]\left[1 + \left(\frac{3f}{4f_*}\right)^2\right]\,.\nn \\ 
    \label{eq:Sigma_Omega_f}
\end{eqnarray}
This definition incorporates the prefactor from Eq.~\eqref{eq:Omega_GW_to_Sh} and simplifies the sum over the $A$ and $E$ channels by noting that both $\mathcal{R}_i(f)$ and $N_i(f)$ are identical for the two.
The second-line approximation follows from the low-frequency form of the noise spectra $N_A(f) = N_E(f)$ [Eq.~\eqref{eq:A_E_noise_spectra}] and the use of the fit to the response function $\mathcal{R}_A(f) = \mathcal{R}_E(f)$ [Eq.~\eqref{eq:A_E_response_functions}].
Note that, since the SNR depends on the detector response and noise spectrum through the ratio $\mathcal{R}_i / N_i$, the factors of $W$ accounting for the additional TDI path lengths in Eqs.~\eqref{eq:A_E_response_functions} and~\eqref{eq:A_E_noise_spectra} cancel exactly.

Given the significant uncertainties surrounding the cosmological background, we restrict our SNR forecasts to the SGWB produced by the three unresolved astrophysical MBHB populations discussed in Sec.~\ref{sec:population_models}. 
To quantify the detectability of the cosmological background, we rely on parameter-error forecasts made using an information-matrix formalism described below.

\subsection{How to Calculate the Information Matrices}
\label{sec:how_to_Fisher}
For the parameter-error forecasts presented in this paper, we model the combined SGWB as the superposition of the cosmological background and the astrophysical background produced by the ensemble of unresolved MBHBs as 
\begin{equation}
\label{eq:combined_background}
    \Omega_{\mathrm{GW}}(f;\,A,\gamma,N_0) = \OmegaCosmo(f;\,A,\gamma)
      + \OmegaAstro(f;\,N_0)\,.
\end{equation}
As mentioned before, $A$ and $\gamma$ are, respectively, the amplitude (defined at the pivot frequency $f_0$) and the spectral index of the cosmological background, while $N_0$ is the present-day ($z\simeq 0$) MBHB merger rate that sets the overall normalization of $\OmegaAstro(f)$.

\textbf{\textit{Fisher information matrix.}}
In order to perform the information matrix analysis, we first need to specify the covariance of the measured phase-shift data set. Assuming that the phase-shift measurements are available in the two independent $A$ and $E$ TDI channels, the mean of the data vanishes and the covariance of the data is diagonal. 

Given the previously detailed assumptions on the detector in Sec.~\ref{sec:exp_spec}, this covariance matrix can be expressed as~\cite{Smith:2019wny}: 
\begin{eqnarray}
    \mathbf{C}_{ab}(f_i) = \frac{1}{2}[\mathcal{S}_a(f_i) + N_a(f_i)]\delta_{ab}\,,
    \label{eq:signal_cov_matrix}
\end{eqnarray}
where $a,\, b\in\{A,\, E\}$, and we have assumed that measurements in different frequency bins are independent. 
Note that the above expression corresponds to the covariance for each measurement made by the detector. 
In other words, if the total dataset for observation time $T_{\rm obs}$ is divided into $1/f_{\rm max}$ intervals, and we assume that the different intervals are statistically independent, then the above matrix represents the covariance in one of the $f_{\rm max} T$ quasi-independent measurements in each frequency bin centered at $f_{i}$. 

Then, in the limit of continuous measurements, for a stationary SGWB, the (inverse) covariance of a set of parameters $\bm{\theta} = \{A, \gamma, N_0\}$ is approximated by the following information matrix:
\begin{eqnarray}
    \label{eq:Fisher-def}
    {\cal F}_{\alpha\beta} &= &\int_{f_{\rm min}}^{f_{\rm max}} \frac{\dd f}{2} \, {\rm \textbf{Tr}}\left[\mathbf{C}^{-1}(f) \, \partial_{\theta_\alpha}\mathbf{C}(f) \, \mathbf{C}^{-1}(f) \, \partial_{\theta_\beta}\mathbf{C}(f)\right]\,, \nn \\
    &= &\frac{T_{\rm obs}}{2} \sum_{i\in \{A,\,E\}} \int_{f_{\rm min}}^{f_{\rm max}} \dd f \, \frac{\partial_{\theta_\alpha}\mathcal{S}_i(f) \, \partial_{\theta_\beta}\mathcal{S}_i(f)}{[N_i(f) + \mathcal{S}_i(f)]^2}\,, \nn \\
    &= &\frac{T_{\rm obs}}{2} \int_{f_{\rm min}}^{f_{\rm max}} \dd f \, \frac{\partial_{\theta_\alpha}\Omega_\GW(f) \, \partial_{\theta_\beta}\Omega_\GW(f)}{[\Sigma_\Omega(f) + \OmegaAstro(f)/\sqrt{2}]^2}\,,
\end{eqnarray}
where, in the third line, we have used the relation in Eq.~(\ref{eq:Omega_GW_to_Sh}) and the effective noise power $\Sigma_{\Omega}(f)$ [Eq.~\eqref{eq:Sigma_Omega_f}] to recast the covariance matrix in Eq.~\eqref{eq:signal_cov_matrix} in terms of $\Omega_\GW(f)$. Here, $\partial_{\theta_\alpha} \equiv \frac{\partial}{\partial \theta_\alpha}$ denotes differentiation with respect to the model parameters $\bm{\theta}$.
Moreover, we include only $\OmegaAstro(f)$ in the denominator, as the unresolved astrophysical background acts as an additional, irreducible noise component that must be distinguished from the cosmological signal.

The $1\sigma$ uncertainties and covariances are obtained from the covariance matrix, ${\cal C}\equiv{\cal F}^{-1}$, as
$\sigma_\alpha=\sqrt{{\cal C}_{\alpha\alpha}}$ and
$\mathrm{Cov}(\theta_\alpha,\theta_\beta)={\cal C}_{\alpha\beta}$.

\textbf{\textit{Analytical derivatives.}}
Because both backgrounds are specified by analytical expressions, the required derivatives can be trivially written as
\begin{align}
    \partial_A\Omega_{\rm GW}(f)      &= \bigl(f/f_0\bigr)^{\gamma}\,, \label{eq:deriv_A} \\
    \partial_\gamma\Omega_{\rm GW}(f) &=\OmegaCosmo(f)\,
                                         \ln{\bigl(f/f_0\bigr)}\,, \\
    \partial_{N_0}\Omega_{\rm GW}(f)  &= \OmegaAstro(f)\,/\,N_0\,.
\end{align}
These expressions make the numerical evaluation of Eq.~\eqref{eq:Fisher-def} fast and robust.

\textbf{\textit{Treatment of the astrophysical background.}} The astrophysical background for each population model depends on the mass-distribution parameters $\alpha$ and $\M_*$, redshift-distribution parameters $\beta$ and $z_0$, and the merger rate $N_0$, as detailed in Sec.~\ref{sec:population_models}.
However, in our parameter error forecasts we choose not to marginalize over the mass- and redshift-distribution parameters.
Instead, we focus only on marginalizing over $N_0$, effectively testing whether an unresolved MBHB background is detectable, while avoiding the vastly larger parameter space associated with a simultaneous inference of population properties.  

One of the primary motivations for this choice is that, in practice, the frequency dependence of $\OmegaAstro(f)$ closely follows the post-Newtonian scaling $f^{2/3}$ throughout the LISA band for all three models considered (Fig.~\ref{fig:Omega_GW_background_allModels}).
In other words, only the overall normalization of $\OmegaAstro(f)$ is determined by the fixed population parameters (set by the models described in Sec.~\ref{sec:population_models}) and by the free parameter $N_0$, which is set to $200\ \mathrm{yr}^{-1}$ in our fiducial analysis.
As detailed in Sec.~\ref{sec:population_models}, variations in the amplitude of $\OmegaAstro(f)$ can be captured entirely by rescaling $N_0$.
Therefore, by simultaneously marginalizing over $\{A, \gamma, N_0\}$ and inverting ${\cal F}$, we automatically account for correlations and potential degeneracies among these three parameters.

\emph{Ultimately, our goal is to quantify the smallest cosmological background amplitude that remains measurable in the presence of the MBHB background, and to assess whether the astrophysical background itself can be detected.}
To this end, we perform a suite of information-matrix forecasts to explore how detectability depends on the primordial parameters $A$ and $\gamma$, the choice of astrophysical population model, and the single-binary detection threshold $\rho_{\mathrm{th}}$.

\begin{table*}[t]
    \centering
    \caption{
        Signal-to-noise ratios for the stochastic differential gravitational-wave energy density $\Omega_{\rm GW}(f)$, assuming a 4-year LISA mission. 
        Each column corresponds to one of the three population models introduced in Tab.~\ref{tab:pop_model_params} and Fig.~\ref{fig:pop_models}, 
        with SNRs computed using Eq.~\eqref{eq:background_SNR}.
        The top row ($\rho_{\rm th} < \infty$) reflects the \textit{total} energy density, including both resolved and unresolved sources. 
        All subsequent rows indicate the energy density of the \textit{background} obtained after removing individually detectable binaries with a detection threshold $\rho_{\rm th}$.
        All SNR values are normalized to a merger rate of $N_0 = 200~\mathrm{yr}^{-1}$; for other merger rates $N_0$, the SNRs scale proportionally as $\left[N_0 / (200~\mathrm{yr}^{-1})\right]$.
    }
    \label{tab:snr_background_models}
    \vspace{6pt}
    \begin{minipage}{\textwidth}
    \centering
    \small
    \renewcommand{\arraystretch}{1.4}
    \begin{tabular}{%
        >{\centering\arraybackslash}p{3.5cm}%
        |>{\centering\arraybackslash}p{3cm}%
        |>{\centering\arraybackslash}p{3cm}%
        |>{\centering\arraybackslash}p{3cm}}
    \hline\hline
    \multirow{2}{=}{\centering \makecell{Individual Binary\\Detection Threshold\\$\rho_{\rm th}$ \\[-10pt]}}  
    & \multicolumn{3}{c}{SNR of $\Omega_{\rm GW}(f) \times \left[N_0 / \left(200~\mathrm{yr}^{-1}\right)\right]$} \\
    \cline{2-4}
      & \makecell[tc]{Model 1\\\texttt{(Light Seed)} \\[2pt]} 
      & \makecell[tc]{Model 2\\\texttt{(Heavy Seed)}} 
      & \makecell[tc]{Model 3\\\texttt{(Ultra-Light Seed)}} \\
    \hline
      $< \infty$ (Total Signal)  & 666 & 14,473 & 43 \\
    \hline
      $\leq 4$   & 2   & 0     & 2 \\
      $\leq 8$   & 12  & 1     & 9 \\
      $\leq 12$  & 27  & 2     & 12 \\
      $\leq 16$  & 46  & 5     & 15 \\
      $\leq 20$  & 68  & 10    & 17 \\
    \hline\hline
    \end{tabular}
    \end{minipage}
\end{table*}

\textbf{\textit{Detection criterion.}}
For parameters that vanish under the null hypothesis---$A$ for the cosmological background and $N_0$ for the astrophysical one---the smallest detectable value at
$n\sigma$ confidence is
\begin{equation}
    A_{\rm min}(n)=n\,\sigma_A\,, \qquad
    N_{0,\,\rm min}(n)=n\,\sigma_{N_0}\,.
\end{equation}
Because $\partial_A\Omega_{\rm GW}(f)$ is independent of $A$,
$\sigma_A$ itself does not depend on the fiducial amplitude. 
So, a single information-matrix evaluation suffices to infer $A_{\rm min}$ at any significance level.  The spectral index $\gamma$ does not satisfy such a simple null hypothesis and is therefore not used as a primary detectability metric.

\textbf{\textit{Implementation.}}
We integrate Eq.~\eqref{eq:Fisher-def} over the effective LISA band $f\in[f_{\min},f_{\max}]$ given in Sec.~\ref{sec:exp_spec} and assume a nominal four-year mission.
Information matrices are computed for each astrophysical-population model introduced in Sec.~\ref{sec:population_models}, for all individual binary detection thresholds considered in Sec.~\ref{sec:astrophysical_background}.
We compute the parameter covariances for a variety of fiducial values of $A$ and $\gamma$, using the ranges specified in Sec.~\ref{sec:primordial_background}.

\subsection{SNR Forecasts}
\label{sec:SNR_forecasts}

Table~\ref{tab:snr_background_models} displays the SNRs for $\OmegaAstro(f)$, computed for a 4-year LISA mission, assuming a merger rate of $N_0 = 200~\mathrm{yr}^{-1}$.
Each column corresponds to one of the three population models introduced in Tab.~\ref{tab:pop_model_params} and Fig.~\ref{fig:pop_models}, with SNRs calculated using the methodology described in Sec.~\ref{sec:how_to_SNR}. 
The top row of Tab.~\ref{tab:snr_background_models} ($\rho_{\rm th} < \infty$) reports the SNR of the \textit{total} energy density, including both resolved and unresolved sources. 
All subsequent rows show the SNR of the \textit{background} after removing individually detectable binaries with (individual) SNRs above the threshold $\rho_{\rm th}$ (see Sec.~\ref{sec:astrophysical_background} for more details on the subtraction). 
Given that the parameter $N_0$ sets the overall normalization of the MBHB population in the LISA band, the SNR forecast for different merger rates can be obtained by rescaling our results by a factor of $N_0 / (200~\mathrm{yr}^{-1})$.

We find that the \textit{total} signal is highest for Model 2 (\texttt{Heavy Seed}), yielding an SNR of $14{,}473$, followed by Model 1 (\texttt{Light Seed}) with $666$, and Model 3 (\texttt{Ultra-Light Seed}) with $43$. 
This ranking is expected from the distribution of binary masses in each scenario: Model 2 predominantly generates heavy binaries merging within the most sensitive part of the LISA band, while Model 3 yields very light binaries that primarily inspiral through the band and merge at higher frequencies beyond LISA's optimal sensitivity.

For the \textit{background-only} signal, a different hierarchy emerges. 
The \texttt{Light Seed} scenario produces the highest background SNRs across all $\rho_{\rm th}$ values, peaking at $68$ for $\rho_{\rm th} = 20$. 
This indicates that a significant fraction of its binaries are unresolved, making it the most likely scenario to generate a detectable astrophysical background in the LISA band. 
Even for conservative choices like $\rho_{\rm th} = 8$, the background SNR reaches $12$, comfortably above the detection threshold. 
Thus, an MBHB population resembling the \texttt{Light Seed} scenario is a promising source of an astrophysical SGWB and can remain detectable even if the true merger rate is appreciably lower than the fiducial value of $N_0 = 200\ {\rm yr}^{-1}$ selected here.

In contrast, while the \texttt{Heavy Seed} scenario produces the highest total signal, the vast majority of its binaries are resolved individually due to their high masses and correspondingly larger (individual) SNRs. 
As a result, the residual background is negligible at low $\rho_{\rm th}$ and remains modest even for higher thresholds; for instance, the background SNR is zero at $\rho_{\rm th} = 4$ and only reaches $2$ at $\rho_{\rm th} = 12$. 
This suggests that this scenario may not generate a detectable background on its own. 
However, it is important to note that its contribution can still affect our ability to constrain a cosmological signal in joint analyses—an issue we examine in Sec.~\ref{sec:Fisher_forecasts}.
Notably, if the time delay between binary formation and merger in this population is significantly longer than in our fiducial model, the effective $N_0$ may be suppressed to the point that the background becomes negligible altogether.

The \texttt{Ultra-Light Seed} scenario occupies an intermediate regime. 
It yields a somehow modest background SNR that slowly grows with $\rho_{\rm th}$, reaching only $17$ at $\rho_{\rm th} = 20$. 
While the majority of its binaries are unresolved, their individual contributions are minuscule due to their substantially lower masses and weaker strain amplitudes. 
This illustrates that a high fraction of unresolved sources does not necessarily imply a strong background---the source masses and their distribution within the LISA band remain critical.

\begin{table*}[t]
    \centering
    \caption{
    Marginalized $1\sigma$ measurement uncertainties on the amplitude $A$ and spectral slope $\gamma$ of a power-law cosmological SGWB, for seven different values of $\gamma$, as measured by LISA. 
    We assume a fiducial amplitude of $A = 10^{-13}$, a reference frequency of $f_0 = 2\ \textrm{mHz}$, and consider only the instrumental sensitivity (i.e., no astrophysical background is included).
    While the uncertainty on $A$, denoted $\sigma_A$, is independent of its fiducial value, the uncertainty on the spectral slope, $\sigma_\gamma$, scales inversely with $A$ as indicated in the column header.
    }
    \label{tab:sigma_A_gamma}
    \vspace{6pt}
    \begin{minipage}{\textwidth}
    \centering
    \small
    \renewcommand{\arraystretch}{1.4}
    \begin{tabular}{%
        >{\centering\arraybackslash}p{3cm}%
        |>{\centering\arraybackslash}p{4cm}%
        |>{\centering\arraybackslash}p{4cm}}
    \hline\hline
    $\gamma$ & $\sigma_A$ & $\sigma_\gamma \left[ A/10^{-13} \right]^{-1}$ \\
    \hline
    $-5.0$ & $1.6 \times 10^{-17}$ & $3.2 \times 10^{-5}$ \\
    $-3.0$ & $1.2 \times 10^{-14}$ & $4.4 \times 10^{-2}$ \\
    $-1.0$ & $4.8 \times 10^{-14}$ & $7.9 \times 10^{-1}$ \\
    $0.0$  & $5.0 \times 10^{-14}$ & $9.6 \times 10^{-1}$ \\
    $1.0$  & $4.7 \times 10^{-14}$ & $6.0 \times 10^{-1}$ \\
    $3.0$  & $6.9 \times 10^{-15}$ & $3.2 \times 10^{-2}$ \\
    $5.0$  & $1.4 \times 10^{-17}$ & $2.6 \times 10^{-5}$ \\
    \hline\hline
    \end{tabular}
    \end{minipage}
\end{table*}

In summary, our forecasts demonstrate that astrophysical backgrounds from MBHBs will be a prominent component of the SGWB in the LISA band. 
Among the models considered, the \texttt{Light Seed} scenario is most conducive to generating a detectable background, even under conservative assumptions. 
The \texttt{Heavy Seed} model, despite producing a strong total signal, is mostly resolvable and thus leaves little residual background. 
It is important to emphasize, however, that these population models are inherently uncertain. 
While our choices reflect current theoretical expectations informed by the literature~\cite{Barausse:2023yrx}, the true underlying population need not correspond to any single formation channel. 
In reality, the merging MBHB population may be a superposition of different formation scenarios. 
For example, a combination of the \LightSeed~and \HeavySeed~channels may potentially result in a stronger or qualitatively different background.
These distinctions become especially relevant when evaluating the prospects for detecting a cosmological SGWB in the presence of an astrophysical background. 
The question of whether, and to what extent, such an astrophysical background will degrade our ability to constrain cosmological signals will be addressed in Sec.~\ref{sec:Fisher_forecasts}.

\subsection{Information-Matrix Forecasts}
\label{sec:Fisher_forecasts}

\begin{figure*}[t]
    \centering
    \includegraphics[width=1.0\textwidth]{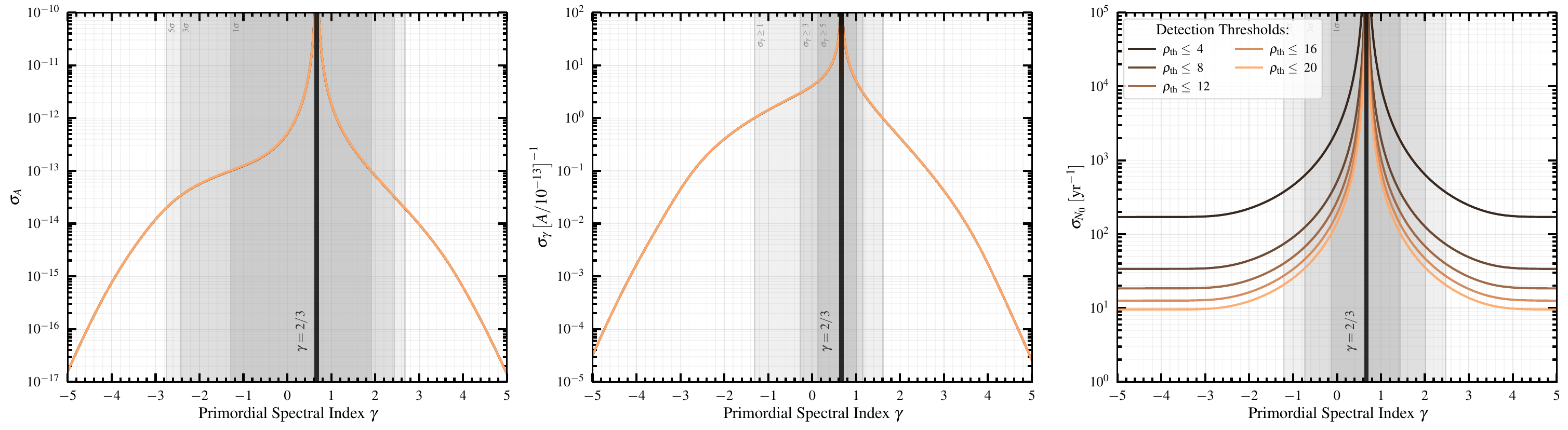} 
    \caption{
    Forecasted $1\sigma$ uncertainties on the amplitude $A$ (left) and spectral index $\gamma$ (center) of the cosmological SGWB, and the local merger rate $N_0$ of the astrophysical SGWB (right), as functions of the spectral index $\gamma$, assuming population Model 1 from Fig.~\ref{fig:pop_models}. 
    The fiducial parameters are set to $N_0 = 200~\mathrm{yr}^{-1}$ and $A = 10^{-13}$. 
    Each curve corresponds to a different individual detection threshold $\rho_{\rm th}$, as indicated in the legend. 
    In the left and right panels, shaded regions indicate the ranges of $\gamma$ where the fractional uncertainty on $A$ and $N_0$ (respectively) exceeds $1\sigma$, $3\sigma$, and $5\sigma$, with darker shades corresponding to lower statistical significance. 
    In the center panel, the shaded regions mark intervals where the absolute uncertainty on $\gamma$ exceeds 1, 3, and 5, respectively. 
    While the fractional uncertainties on $A$ and $N_0$ are independent of the fiducial value of $A$, the absolute uncertainty $\sigma_\gamma$ scales inversely with it, as shown. 
    The vertical line at $\gamma = 2/3$ denotes the spectral index at which the cosmological and astrophysical backgrounds become degenerate in spectral shape, leading to a sharp increase in parameter uncertainties. 
    Results for Models 2 and 3 are qualitatively similar.
    }
    \label{fig:sigmaA_gamma_N0}
\end{figure*}

\begin{figure}[h]
    \centering
    \includegraphics[width=0.5\textwidth]{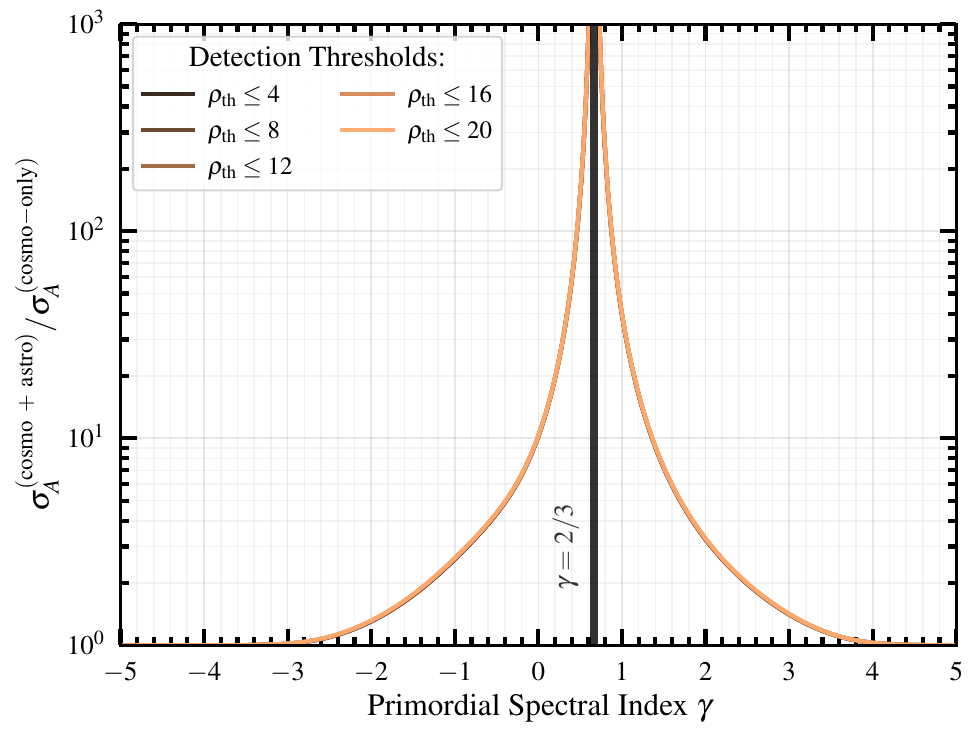} 
    \caption{
    Ratio of the $1\sigma$ uncertainty on the amplitude $A$ of a cosmological SGWB obtained from a joint analysis including both cosmological and astrophysical components to that from a cosmological-only analysis, as a function of the spectral index $\gamma$. 
    Each curve corresponds to a different detection threshold $\rho_{\rm th}$ for individually resolved binaries, as indicated in the legend. 
    This figure quantifies the impact of the unresolved astrophysical background on our ability to detect and characterize a cosmological SGWB. 
    The degradation in sensitivity peaks near $\gamma = 2/3$, where the spectral shapes of the two components become nearly degenerate, significantly inflating the uncertainty on $A$. 
    Away from this degeneracy, the influence of the astrophysical background becomes strongly $\gamma$-dependent and is negligible for $\gamma \lesssim -3$ and $\gamma \gtrsim 3$.
    }
    \label{fig:sigmaA_prim_vs_astro}
\end{figure}

To understand how the astrophysical background impacts the detectability and characterization of a cosmological SGWB, we perform a series of information-matrix forecasts. 
We begin by analyzing a baseline case that includes only the cosmological background, before incorporating the astrophysical background sourced by unresolved MBHBs.
The methodology underlying these calculations is detailed in Sec.~\ref{sec:how_to_Fisher}.

\textbf{\textit{Baseline case: cosmological-background only.}} In the baseline scenario, we assume the SGWB is solely of cosmological origin, described by a power law of amplitude $A$ and spectral index $\gamma$, and neglect any astrophysical contribution. 
This allows us to assess LISA's ability to measure $A$ and $\gamma$ in the presence of instrumental noise alone. 
Table~\ref{tab:sigma_A_gamma} presents the resulting $1\sigma$ marginalized uncertainties $\sigma_A$ and $\sigma_\gamma$ for seven fiducial values of $\gamma$ ranging from $-5$ to $5$. The amplitude is fixed to $A=10^{-13}$, and we adopt a pivot frequency $f_0 = 2\ \mathrm{mHz}$ and a 4-year mission duration.

Given the derivatives in Sec.~\ref{sec:how_to_Fisher}, it is straightforward to derive that $\sigma_A$ is independent of the fiducial values of $A$, whereas $\sigma_\gamma$ scales inversely with it.
The uncertainties are largest near $\gamma = 0$ and decrease symmetrically toward large positive or negative spectral indices. 
This behavior reflects the nature of power-law backgrounds: flatter spectra distribute power more uniformly across frequency, reducing contrast with the frequency-dependent instrumental sensitivity.
For instance, at $\gamma = 0$, we find $\sigma_A = 5.0 \times 10^{-14}$, implying that a $3\sigma$ ($5\sigma$) detection will require $A \gtrsim 1.5\times10^{-13}$ ($2.5\times10^{-13}$).
Steep spectra, by contrast, are easier to distinguish. 
For example, for $\gamma = \pm 3$, $\sigma_A$ is up to an order of magnitude lower. 
Overall, these results demonstrate that, in the absence of astrophysical backgrounds, LISA will be sensitive to cosmological backgrounds with amplitudes $A \gtrsim 10^{-13}$ across a wide range of spectral indices, with improved sensitivity for steeper spectra (larger $|\gamma|$).

\textbf{\textit{Combined analysis: cosmological + astrophysical backgrounds.}} We now turn to the full three-parameter information-matrix analysis that incorporates both the cosmological and astrophysical components, as defined in Eq.~\eqref{eq:combined_background}. 
The methodology underlying these forecasts is detailed in Sec.~\ref{sec:how_to_Fisher}. 

Figure~\ref{fig:sigmaA_gamma_N0} shows the $1\sigma$ uncertainties on the cosmological amplitude $A$ (left panel), the spectral index $\gamma$ (center), and the astrophysical merger rate $N_0$ (right), as functions of the fiducial spectral index $\gamma$. 
Results are shown for population Model~1 (\texttt{Light Seed}), with fiducial parameters $N_0 = 200~\mathrm{yr}^{-1}$ and $A = 10^{-13}$. 
Each curve corresponds to a different single-binary detection threshold $\rho_\mathrm{th}$.

In the left and right panels, the shaded regions indicate intervals where the fractional uncertainty on $A$ and $N_0$, respectively, exceeds $1\sigma$, $3\sigma$, and $5\sigma$, with darker shading corresponding to lower statistical significance. 
In the center panel, the shaded bands denote regions where the absolute uncertainty on $\gamma$ exceeds 1, 3, or 5. 
These visual guides help identify the ranges of $\gamma$ where each parameter becomes more or less measurable.

While $\sigma_A$ and $\sigma_{N_0}$ are independent of the fiducial amplitude $A$, $\sigma_\gamma$ scales inversely with $A$, as indicated by the axis labels. 
The vertical black line at $\gamma = 2/3$ marks the region where the spectral shapes of the cosmological and astrophysical backgrounds become nearly degenerate, leading to a degradation in parameter constraints.
Although the figure shows results for Model~1, we observe qualitatively similar trends for the other population models considered.

Compared to the baseline case, the inclusion of the astrophysical background significantly inflates $\sigma_A$ specifically in regions where the two spectral components scale similarly with frequency. 
This overlap is most problematic near $\gamma = 2/3$, where the cosmological signal mimics the $f^{2/3}$ slope of the astrophysical background in LISA's most sensitive frequency range (near $\sim$mHz), leading to a pronounced degeneracy and increased uncertainty on $A$.
For instance, at $\gamma = 0$ (flat spectrum), we find $\sigma_A \approx 5\times10^{-13}$, implying that a detectable cosmological amplitude must exceed $A \gtrsim 1.5\times10^{-12}$ ($2.5\times10^{-12}$) for $3\sigma$ ($5\sigma$) detection, \textit{an order-of-magnitude increase relative to the cosmological-background-only scenario}.
The impact is even larger at $\gamma = 1$, with a degradation factor of $\sim 40$.
By contrast, for $\gamma \lesssim -3$ or $\gamma \gtrsim 3$, the degradation is negligible.

Figure~\ref{fig:sigmaA_prim_vs_astro} quantifies the degree of degradation induced by the existence of the astrophysical background by displaying the ratio of $\sigma_A$ obtained in the combined (cosmo + astro) scenario to that from the cosmological-background-only analysis (cosmo-only), as a function of $\gamma$. 
As expected, the largest degradation occurs near $\gamma = 2/3$. 
\textit{Vitally, this degeneracy is rooted in spectral shape, not amplitude:} changes in $\rho_\mathrm{th}$ or $N_0$ affect the strength of the astrophysical background and thus the detectability of $N_0$, but they have limited impact on the ability to distinguish two components with nearly identical spectral slopes.

The uncertainties on $\gamma$ (center panel of Fig.~\ref{fig:sigmaA_gamma_N0}) follow trends similar to those on $A$, increasing sharply near $\gamma = 2/3$. 
Nevertheless, for most values of $\gamma$, $\sigma_\gamma < 1$, indicating that the spectral index of the cosmological background can generally be constrained with good precision. 
Unlike with $\sigma_A$, these results scale inversely with $A$, as indicated on the middle panel axis label.

In contrast, the uncertainty on $N_0$ is most sensitive to $\rho_\mathrm{th}$ and only weakly dependent on $\gamma$ outside the range $-1 < \gamma < 2$. 
When the cosmological background rises or falls steeply, its spectral shape diverges significantly from the $f^{2/3}$ astrophysical background, enabling efficient disentanglement. 
The tightest constraints on $N_0$ arise for high values of $\rho_\mathrm{th}$, where the residual astrophysical background is strongest. 
For $\rho_\mathrm{th} = 4$, however, the background is too weak to yield meaningful constraints: even for $N_0 = 200~\mathrm{yr}^{-1}$, the uncertainty $\sigma_{N_0}$ remains large. 
If the true merger rate is lower, e.g., $N_0 \lesssim 20~\mathrm{yr}^{-1}$, $\sigma_{N_0}$ exceeds the fiducial value across all $\gamma$, rendering the astrophysical background effectively undetectable.

\paragraph*{\textbf{Alternate case: fixed spectral index $\gamma = 0$.}} 
While we focus here on a general cosmological background characterized by both amplitude and spectral index $\{A,\, \gamma\}$, some early-Universe scenarios, such as certain models of inflation or string-gas cosmology, predict a nearly flat gravitational-wave spectrum. 
In such cases, one may fix $\gamma = 0$ and perform a reduced two-parameter information-matrix analysis in $\{A,\, N_0\}$, avoiding marginalization over $\gamma$. 
Results for this special case, including the impact of varying the single-binary detection threshold $\rho_{\rm th}$ and local merger rate $N_0$, are presented in Appendix~\ref{app:flat_spectrum_forecasts}. 
These forecasts provide useful insight into scenarios where the cosmological background shape is known a priori.

\textbf{\textit{Astrophysical complexity.}} 
In addition to the stochastic background from unresolved MBHBs, LISA will be sensitive to other overlapping astrophysical contributions, including those from EMRIs and stellar-mass binary black holes.
These sources may also produce overlapping stochastic signals, with spectral shapes that are partially degenerate with each other and with the cosmological background.
For instance, the stellar-mass binary black hole background is expected to follow a characteristic $f^{2/3}$ scaling in the LISA band~\cite{Babak:2023lro}, similar to both MBHBs and some cosmological models, while EMRI-induced backgrounds may exhibit comparable degeneracies depending on their population properties~\cite{Bonetti:2020jku}. 
The presence of an unresolved MBHB background will therefore hinder not only the detection of a cosmological SGWB, but also the identification and characterization of these additional astrophysical components. 
Therefore, accurate modeling and subtraction of the MBHB contribution is essential for both fundamental cosmology and for inferences on the EMRI and stellar-mass black hole populations.

Disentangling these signals requires a multi-faceted approach: leveraging any available spectral differences, constraining unresolved backgrounds through the properties of resolved events (both in the LISA band and from ground-based detectors), and incorporating alternative statistical techniques. 
A particularly promising avenue for disentangling overlapping backgrounds lies in quantifying higher-order statistical properties or the non-Gaussianity of the SGWB.
For example, the stellar-mass binary background, expected to originate from a large number of weak, early-inspiral sources, results in a signal that is more Gaussian than that of the MBHB background.
Therefore, the use of alternative statistical techniques will likely be pivotal in the pursuit of separating overlapping SGWB components in the LISA band.

\textbf{\textit{Implications.}} \textit{Taken together, our forecasts demonstrate that astrophysical foregrounds sourced by unresolved MBHBs will substantially impair the detectability and characterization of a cosmological SGWB, even under conservative assumptions about source subtraction.}
This degradation is most pronounced near $\gamma = 2/3$, where the spectral slopes of the cosmological and astrophysical components become nearly degenerate within the most sensitive region of the LISA band. 
However, we find that significant impact persists well beyond this narrow degeneracy point: for example, even flat ($\gamma = 0$) or mildly blue ($\gamma = 1$) cosmological spectra (both of which are plausible in various early-Universe scenarios) require order-of-magnitude larger amplitudes to remain detectable in the presence of MBHB confusion noise. 
These results hold even when assuming optimistically low detection thresholds ($\rho_{\rm th} \leq 4$), where one might expect most MBHBs to be subtracted. 
As shown in Sec.~\ref{sec:SNR_forecasts}, however, such low thresholds still leave behind substantial residual backgrounds in plausible population scenarios, particularly in \texttt{Light Seed}-like models, which are likely to generate detectable astrophysical backgrounds. 
Importantly, this confusion noise persists even for local merger rates $N_0$ below our fiducial value, further reinforcing the need to account for it when interpreting LISA’s sensitivity to cosmological signals.

\begin{figure*}[t]
    \centering
    \includegraphics[width=1.0\textwidth]{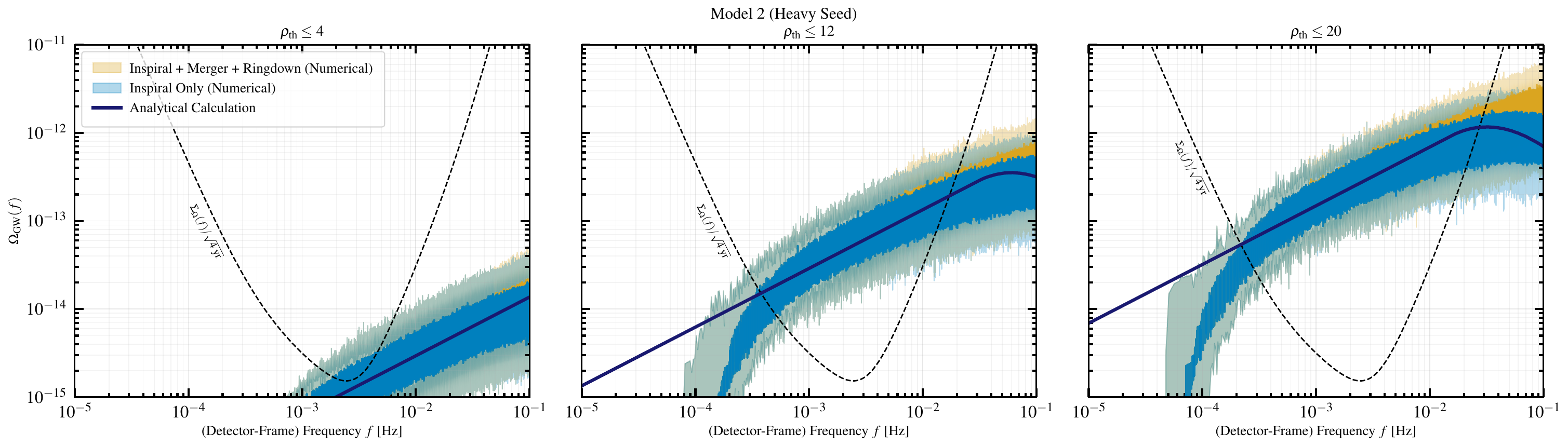} 
    \caption{
    Numerical estimation of the SED of the astrophysical SGWB from unresolved MBHBs using the Model 2 (Heavy Seed) population.  
    Each panel corresponds to a different individual-binary detection threshold, $\rho_{\mathrm{th}} = 4,\,12,\,20$ (left to right), where only binaries with SNR below the threshold contribute to the background.
    High-opacity bands show the median $\Omega_{\rm GW}(f)$ across 50 Monte Carlo realizations, while low-opacity bands represent the corresponding 85\% confidence intervals. 
    Each realization is computed using either the full \texttt{IMRPhenomX} waveform (golden yellow) or the inspiral-only waveform truncated at the ISCO (light blue). 
    The solid navy line shows the analytical result for the same population, and agrees well with the inspiral-only numerical calculation.
    The dashed black curve shows the nominal LISA sensitivity $\Sigma_\Omega(f)$, scaled down to account for an observation period of $T_{\rm obs} = 4$ years.  
    The sharp cutoff and deviation from power-law behavior at low frequencies arise because our numerical calculation of $\OmegaGW(f)$ only tracks MBHBs that merge within the 4-year observation period of a LISA-like detector. 
    This leads to a suppression of the signal from low-$\M$ binaries in their early inspiral phase that are expected to merge after the observation ends.
    While the merger and ringdown phases slightly increase the background at higher $\rho_{\mathrm{th}}$, where more massive and shorter-lived binaries remain unresolved, their overall contribution remains small.  
    Models 1 and 3 show qualitatively similar behavior.
    }
    \label{fig:Omega_GW_background_numerical_Model2}
\end{figure*}

Conversely, the astrophysical background itself can often be distinguished from a cosmological component, but this depends on the spectral index $\gamma$, the resolving threshold $\rho_{\rm th}$, and the underlying merger rate $N_0$. 
For steep or shallow primordial spectra ($\gamma \lesssim -3$ or $\gtrsim 3$), disentanglement is straightforward due to the stark contrast in spectral shapes. 
In more ambiguous cases, especially within $-1 < \gamma < 2$, we find that constraining the astrophysical component requires sufficiently high $\rho_{\rm th}$ and merger rates; for instance, if $N_0 \lesssim 20~\mathrm{yr}^{-1}$, the background may evade detection altogether.
Therefore, \textit{our results underscore that the astrophysical SGWB is not merely a nuisance but a dynamically relevant component of the total background, one that must be carefully modeled and marginalized over in any joint analysis aimed at probing fundamental physics with LISA.}

\subsection{The Numerical Check}
\label{sec:the_numerical_check_forecasts}

\begin{table*}[t]
    \centering
    \caption{
    Comparison between the analytically predicted and numerically estimated SNRs of the unresolved astrophysical background $\OmegaAstro(f)$ from MBHBs, for various individual-binary detection thresholds $\rho_{\rm th}$. 
    All results assume the Model~2 (\texttt{Heavy Seed}) population with a local merger rate of $N_0 = 200~\mathrm{yr}^{-1}$ and a 4-year LISA observation period. 
    The left and middle columns show the numerically calculated SNRs using the full (inspiral + merger + ringdown) waveform and the inspiral-only waveform truncated at ISCO, respectively. 
    Each entry denotes the median SNR over 50 Monte Carlo realizations, with the 85\% confidence interval indicated. 
    The rightmost column shows the analytical prediction from Eq.~\eqref{eqn:final_astro_background}. 
    All approaches yield consistent results, validating the analytical formalism developed. 
    At high $\rho_{\rm th}$, the inclusion of merger and ringdown phases slightly increases the SNR, but this enhancement remains modest. 
    Qualitatively similar agreement is observed for Models~1 and~3, with even closer correspondence between numerical and analytical estimates.
    Note that SNRs reported here scale with the local merger rate as $N_0 / \left(200~\mathrm{yr}^{-1}\right)$.
    }
    \label{tab:snr_thresholds}
    \vspace{6pt}
    \begin{minipage}{\textwidth}
    \centering
    \small
    \renewcommand{\arraystretch}{1.4}
    \begin{tabular}{%
        >{\centering\arraybackslash}p{3.5cm}%
        |>{\centering\arraybackslash}p{3cm}%
        |>{\centering\arraybackslash}p{3cm}%
        |>{\centering\arraybackslash}p{3cm}}
    \hline\hline
    \multirow{2}{=}{\centering \makecell{Individual Binary\\Detection Threshold\\$\rho_{\rm th}$ \\[-10pt]}}  
    & \multicolumn{3}{c}{SNR of $\Omega_{\rm GW}(f) \times \left[N_0 /\left(200~\mathrm{yr}^{-1}\right)\right]$} \\
    \cline{2-4}
        & \makecell[tc]{Inspiral + Merger +\\ Ringdown (Num.)} 
        & \makecell[tc]{Inspiral Only\\ (Num.)} 
        & \makecell[tc]{Analytical\\\scriptsize [Eq.~\eqref{eqn:final_astro_background}]} \\
    \hline
    $\leq 4$  & $0.0^{+0.1}$ & $0.0^{+0.1}$ & $0$ \\
    $\leq 8$  & $0.6^{+1.7}_{-0.2}$ & $0.6^{+1.7}_{-0.2}$ & $1$ \\
    $\leq 12$ & $2.3^{+5.6}_{-0.9}$ & $2.3^{+5.6}_{-0.9}$ & $2$ \\
    $\leq 16$ & $5.7^{+13.2}_{-2.2}$ & $5.7^{+13.1}_{-2.2}$ & $5$ \\
    $\leq 20$ & $11.3^{+26.2}_{-4.5}$ & $11.0^{+24.6}_{-4.4}$ & $10$ \\
    \hline\hline
    \end{tabular}
    \end{minipage}
\end{table*}

The forecasts presented in Sec.~\ref{sec:Fisher_forecasts} rely on the analytical model introduced in Secs.~\ref{sec:spectral_density} and~\ref{sec:astrophysical_background}.
This model estimates the astrophysical SGWB assuming a sky-averaged, stationary MBHB population characterized solely by chirp mass $\M$ and redshift $z$, with equal-mass ($q = 1$) and non-spinning binaries.
More critically, the model further assumes that the predicted $\OmegaAstro(f)$ can be computed to good approximation using inspiral-only waveforms truncated at the ISCO.

While this formulation captures the leading-order behavior of the unresolved background and allows for efficient information-matrix forecasting, its validity, especially near LISA’s peak sensitivity, must be tested against more complete waveform models.
Therefore, our goal is to evaluate the accuracy of this analytic framework by comparing it to fully numerical estimates based on inspiral-merger-ringdown waveforms, which account for the full time-domain evolution of each binary. 
This numerical check probes not only the ISCO truncation but also the fidelity of the assumed population and spectral averaging procedures in reproducing the actual SGWB from a Monte Carlo sampled population of MBHBs.

To this end, we conduct a fully numerical calculation of the background using Monte Carlo realizations of the underlying population. 
For each realization, we sample binaries' $\M $ and $z$ from the merger rate distribution described in Eq.~\eqref{eq:merger_rate_definition} (for all the three population models described in Sec.~\ref{sec:population_models}).
Matching the simplifying assumptions we made in characterizing the analytical model for $\dd E_\GW/\dd \ln f_r$ [Eq.~\eqref{eq:individual_MBHB_SED}], we assume the binaries have circular orbits, are equal mass and nonspinning.
We randomize their right ascension, declination, inclination, polarization angle, and phase of coalescence and uniformly distribute the coalescence times over the 4-year LISA mission duration. 

The waveform for each source is generated using the \texttt{IMRPhenomX} model, which includes inspiral, merger, and ringdown phases accounting only for the dominant $(\ell, m) = (2, 2)$ mode~\cite{Pratten:2020fqn}.
Given the Fourier transforms of the plus and cross polarizations of the gravitational wave strain at the detector, $h_+(f)$ and $h_\times(f)$, we compute the total one-sided strain power spectral density $S_h(f)$ [Eq.~\ref{eq:PS_definition}] for the entire population, and use Eq.~\eqref{eq:Omega_GW_to_Sh} to convert this into a numerical estimate of $\Omega_{\rm GW}(f)$. 
This process is then repeated for 50 independent realizations.
We also generate inspiral-only versions of the same waveforms, truncated at the redshifted ISCO frequency for each binary, enabling a direct numerical comparison to the analytic formalism.

Figure~\ref{fig:Omega_GW_background_numerical_Model2} presents the astrophysical background $\OmegaAstro(f)$ for the \texttt{Heavy Seed} population model, for three representative values of the single-binary detection threshold: $\rho_{\rm th} = 4$, $12$, and $20$ (left, center, and right panels, respectively). 
In each panel, the solid navy line shows the analytical result, while the golden yellow and light blue bands represent the IMR and inspiral-only numerical realizations, respectively. 
High-opacity curves denote the median $\OmegaAstro(f)$, while the lighter bands indicate the corresponding 85\% confidence intervals. 
The dashed black line shows the nominal LISA sensitivity. 
The excellent agreement between the analytical model and the inspiral-only numerical results confirms that our ISCO-truncated treatment accurately captures both the amplitude and spectral shape of the unresolved background.

This agreement is further corroborated in Tab.~\ref{tab:snr_thresholds}, which compares the SNRs computed from the analytical expression and from both numerical approaches---one including only the inspiral phase and the other incorporating IMR waveform.
Each entry represents the median SNR across 50 Monte Carlo realizations, with the 85\% confidence interval reported.
All results assume the \texttt{Heavy Seed} population model with $N_0 = 200~\mathrm{yr}^{-1}$ and a 4-year LISA mission.
The near-perfect agreement across the board confirms that the analytic method faithfully captures the average background amplitude under realistic assumptions, even in the presence of merger contributions. 

At higher frequencies, the inclusion of merger and ringdown phases leads to a modest increase in the background, particularly for larger values of $\rho_{\rm th}$. 
This enhancement originates from lighter binaries that are not individually detectable, merging near the edge of the LISA band. 
Since these MBHBs merge at frequencies slightly higher than their $f_{r,{\rm ISCO}}$, the yellow curve in each panel appears from behind the $f_{r,{\rm ISCO}}$-truncated numerical results and deviates later.
However, this effect remains subdominant; the associated deviations, which become visually noticeable above $\sim 10^{-2}$ Hz for $\rho_{\rm th} = 12$ and $20$, lie largely within the noise-dominated regime of the detector and do not affect detectability.

In the low-frequency regime, deviations from the expected $f^{2/3}$ scaling appear below $\sim 10^{-3}$ Hz. 
These features arise because our numerical simulations only track MBHBs that merge within the 4-year observation window, neglecting signals from systems that merge outside the $t \in [0, 4]\ \mathrm{yr}$ interval. 
In other words, the deviation is not physically meaningful but instead reflects a limitation of the numerical population and waveform synthesis.
This choice is motivated by both computational efficiency and physical considerations: binaries that merge after the mission ends are in their early inspiral phase during observation. 
These early inspirals are extremely weak and nearly monochromatic, so their omission has little impact on the overall astrophysical background detectability, except for the slight downturn at the lowest frequencies.

The \texttt{Heavy Seed} model is chosen for this comparison because, on average, it produces more massive binaries, and hence it presents the largest potential contribution from merger and ringdown. The fact that we observe good agreement in this case implies that our analytical formalism is even more robust for other models (e.g., \texttt{Light Seed}), which generate lower-mass binaries with longer inspirals. We confirm this by verifying that Models 1 and 3 produce qualitatively similar spectra with even tighter agreement between analytic and numerical estimates.

\paragraph*{\textbf{Limitations for resolved binaries.}}
While our analytic prescription accurately captures the unresolved background, it is not designed to model $\Omega_{\rm GW}(f)$ from individually resolved binaries.
These sources, particularly those merging in band near the peak sensitivity of LISA, exhibit strong time-domain structure and non-Gaussian features associated with the merger and ringdown phases.
Since our analytical framework truncates waveforms at the ISCO and assumes smooth spectral averaging, it omits these sharp features and the associated energy contribution.
As such, the approach is reliable for modeling the isotropic stochastic background, but cannot be reliably applied to compute $\Omega_{\rm GW}(f)$ for resolved events or for anisotropic, transient signals.

\paragraph*{\textbf{On statistical limitations of $\Omega_{\rm GW}(f)$.}}
We also emphasize that $\Omega_{\rm GW}(f)$, as a two-point correlation function in frequency space, is intrinsically insensitive to higher-order structure in the GW signal.
If the underlying SGWB contains significant non-Gaussianity, whether sourced by strong individual events, clustering of sources, or foreground anisotropies, this will not be captured by $\Omega_{\rm GW}(f)$ alone.
Alternative probes such as the bispectrum, higher-order moments, or time-frequency domain statistics will be required to detect such deviations from Gaussianity in the gravitational-wave background.
We defer a detailed investigation of such complex features to future work.


\section{Conclusions}
\label{sec:conclusions}

Disentangling the rich astrophysical structure of the SGWB is essential to uncover potential cosmological or primordial signals within LISA’s sensitivity band and to advance our understanding of the early Universe. 
While cosmological sources of the SGWB (including inflation, phase transitions, and relic topological defects) may leave detectable imprints in the SGWB, their discovery critically depends on accurately modeling and subtracting foregrounds from late-time astrophysical sources.
In the $\sim \textrm{mHz}$ band probed by LISA, one of the most significant yet previously underappreciated contributors to this foreground is the unresolved MBHB-induced SGWB.
Though many MBHBs will be individually resolvable, a substantial population, particularly at lower masses and higher redshifts, will remain below the detection thresholds, forming a potentially non-Gaussian, frequency-dependent background that convolutes the search for cosmological signals.
Understanding the statistical and spectral properties of this background, therefore, is a prerequisite for realizing the full scientific potential of LISA.

In this work, we conducted a detailed investigation into the astrophysical SGWB produced by unresolved MBHBs and its impact on the detection of the cosmological background within the LISA band.
Our analysis provides the first comprehensive quantification of how this background influences the detectability of the cosmological SGWB, vitally determining the smallest detectable cosmological background amplitude in the presence of realistic MBHB-induced SGWBs.

For this, we developed an analytical framework that computes the $\OmegaGW(f)$ from MBHB populations across a wide range of mass and redshift distributions, including the effects of binary mergers occurring within or near the LISA band. 
Our formalism includes a detectability criterion for individual binaries and is validated against Monte Carlo simulations using full IMR waveforms. 
This framework enables efficient and physically grounded forecasts of the unresolved MBHB background, serving both as a tool for astrophysical inference and as a crucial input for cosmological SGWB searches.

Given the above framework, we then examine three representative MBHB formation scenarios---the \texttt{Light Seed}, \texttt{Heavy Seed}, and \texttt{Ultra-Light Seed}---thus capturing possible diversity in the resulting astrophysical SGWB. 
Our results show that, while the \texttt{Heavy Seed} model yields the largest total (resolved + unresolved) $\OmegaGW(f)$, most of its sources are individually resolvable, leaving a residual background with ${\rm SNR} \approx 2$ for a (individual) detection threshold $\rho_{\rm th} = 12$. 
In contrast, the \texttt{Light Seed} scenario produces the strongest unresolved background, despite its lower total energy output. Under the same detection threshold, it yields ${\rm SNR} \approx 27$, making it the most significant astrophysical contaminant among the cases considered.
The \texttt{Ultra-Light Seed} case, composed of low-mass binaries, leads to a comparatively weak unresolved signal (${\rm SNR} \approx 10$), illustrating that a higher fraction of unresolved binaries does not necessarily imply a stronger background. 
While these models were studied individually, we stress that real astrophysical foregrounds may result from a combination of such channels, potentially amplifying their cumulative impact.

To assess how this foreground affects cosmological SGWB detection, we performed a joint information-matrix analysis across a broad range of cosmological amplitudes and spectral indices. 
We quantified the degradation in detectability by determining the minimum measurable amplitude $A_{\rm min}$ for each case. 
We find that unresolved MBHBs significantly hinder cosmological SGWB searches across a wide range of spectral slopes, particularly for $-0.5 \lesssim \gamma \lesssim 2$, where the presence of the MBHB-induced foreground leads to a degradation by a factor of 5 or more in the measurement of cosmological background parameters.
For a scale-invariant cosmological background ($\gamma = 0$), the minimum detectable amplitude increases from $1.5 \times 10^{-13}$ (in the absence of astrophysical foregrounds) to $1.5 \times 10^{-12}$ for a $3\sigma$ detection. 
The deterioration is even more severe near $\gamma = 1$ (for $f_0 = 2\,{\rm mHz}$), with $A_{\rm min}$ worsening by a factor of $\sim 40$, reaching $5.4 \times 10^{-12}$. The effect remains notable at $\gamma = 3$, where $A_{\rm min} \approx 3.0 \times 10^{-14}$.

Our analysis also underscores that more optimistic scenarios for individual binary subtraction, corresponding to low detection thresholds of $\rho_{\rm th} \leq 4$–$8$, still result in substantial residual SGWB power for plausible MBHB populations. 
These residuals limit our ability to isolate and characterize cosmological signals, making accurate modeling of the unresolved astrophysical background indispensable for future detection efforts.

Crucially, this framework not only informs strategies for disentangling astrophysical and cosmological SGWBs but also provides a unique observational window into MBHB populations that are otherwise inaccessible. 
In particular, binaries that merge between the LISA and ground-based detector bands, or those too light to be detected by pulsar timing arrays, may contribute significantly to the LISA-band SGWB. 
The unresolved background thus provides a new avenue to probe the demographics of massive black holes that lie beyond the reach of individual detection.

Looking ahead, fully unlocking LISA’s potential to probe fundamental physics will require sustained theoretical, computational, and observational progress in modeling the astrophysical background. 
In particular, narrowing uncertainties in the demographics of MBHBs, through deep galaxy surveys, PTAs, and eventually LISA's own detections, will be crucial for disentangling their contribution to the SGWB. 
Equally essential is the development of robust, scalable data analysis strategies capable of jointly extracting astrophysical and cosmological information from the observed GW signal. 
The framework introduced in this work provides a physically grounded and computationally efficient approach for modeling the unresolved MBHB background.
We also note, however, that $\OmegaGW(f)$ captures only the two-point correlation structure of the signal.
As such, it remains blind to higher-order statistics that may encode rich information about the underlying population, such as the (time-domain) non-Gaussianity induced by the so-called ``popcorn'' nature of the background, spatial clustering, or anisotropic source distributions.
Unveiling these features will require moving beyond two-point statistics to incorporate more effective probes such as the bispectrum, time-frequency domain techniques, or more. 
Developing such complementary diagnostics will be essential to fully characterize the SGWB and to realize the discovery potential of LISA in the presence of complex foregrounds.


\acknowledgements

We thank Emanuele Berti, Daniel D’Orazio, Francesco Iacovelli, Luca Reali, and Benjamin Wandelt for useful discussions. 
We are grateful to Sylvain Marsat for his guidance on the use of the \texttt{lisabeta} code and his assistance with the dataset generation used in the numerical validation of our analytical model.
This work was supported at JHU by NSF Grant No.~2412361, NASA ATP Grant No.~80NSSC24K1226, the Guggenheim Foundation, and the Templeton Foundation.
M.K.~thanks the Center for Computational Astrophysics at the Flatiron Institute and the Institute for Advanced Study for their hospitality. 
This work was partly carried out at the Advanced Research Computing at Hopkins (ARCH) core facility (\url{rockfish.jhu.edu}), which is supported by the NSF Grant No.~OAC-1920103.

\label{app:Mzparameters}
\begin{figure*}
    \centering
    \includegraphics[width=1.0\textwidth]{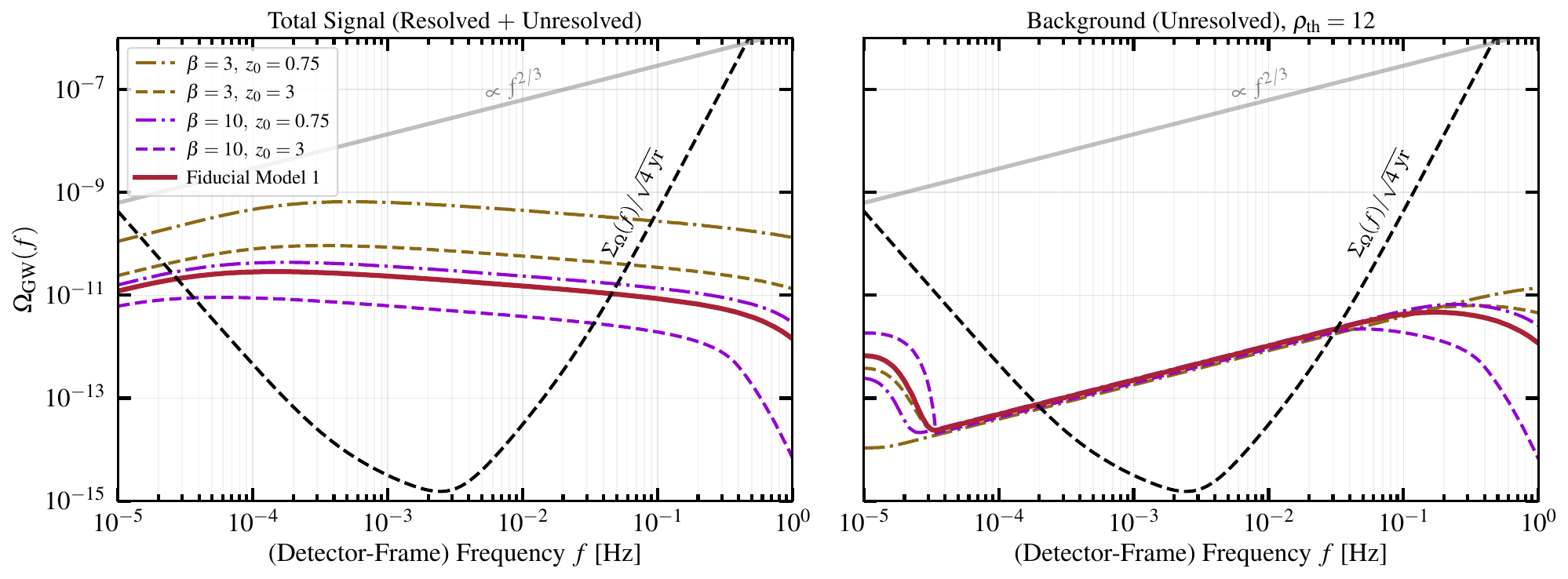} 
    \caption{
    Impact of varying the redshift distribution parameters $\beta$ and $z_0$ on the \textit{total} $\OmegaGW(f)$ (left) and the unresolved \textit{background} $\OmegaAstro(f)$ (right), both sourced by inspiraling and merging MBHBs in the LISA band. 
    The mass distribution follows the \LightSeed~scenario (Tab.~\ref{tab:pop_model_params}). 
    Yellow (purple) curves correspond to $\beta = 3$ ($\beta = 10$), while dot-dashed (dashed) lines correspond to $z_0 = 0.75$ ($z_0 = 3$). 
    Results from the fiducial \LightSeed~model are also displayed (solid red) to provide an additional benchmark for reference. 
    Finally, both panels also include the characteristic $\propto f^{2/3}$ spectrum (gray, expected from purely inspiraling sources), and the LISA sensitivity curve (black, scaled by $T_{\rm obs}^{-1/2}$).
    Changes in the redshift distribution have a mild effect on both the \textit{total} $\OmegaGW(f)$ and the background $\OmegaAstro(f)$, though the impact on $\OmegaAstro(f)$ is more visible at frequencies where the signal deviates from $\propto f^{2/3}$ due to MBHB mergers.
    }
    \label{fig:OmegaAstro_z_dependence}
\end{figure*} 


\appendix

\section{Effect of Varying the Redshift Distribution Parameters}
\label{sec:varying_redshift_params}
Given that the three population models considered in this work (Sec.~\ref{sec:population_models}) are parametrized by roughly similar redshift distributions, we dedicate this section to exploring the sensitivity of the predicted $\OmegaAstro(f)$ to the redshift distribution of MBHB sources.

For our analysis, we choose the \LightSeed~model to characterize the mass distribution and focus on varying the parameters $\beta$ and $z_0$.
Given this fixed $\M$-distribution, we analytically compute both the \textit{total} (resolved + unresolved) $\OmegaGW(f)$ and the unresolved $\OmegaAstro(f)$ for four different pairs of fiducial values assigned to $\beta$ and $z_0$---each unique pair is chosen by selecting $\beta \in \{3,\, 10\}$ and $z_0 \in \{0.75,\, 3\}$.
The pair $\beta = 3$ and $z_0 = 0.75$ corresponds to an MBHB redshift distribution centered around $ z \lesssim 3 $, while $\beta = 10$ and $z_0 = 3$ yield a distribution whose support extends to $ z \gtrsim 12 $, where it becomes most prominent.
In other words, smaller values of $\beta$ and $z_0$ result in a source distribution that is more concentrated at low redshifts.

The left panel of Fig.~\ref{fig:OmegaAstro_z_dependence} shows the predicted SED of the \textit{total} astrophysical signal from the combined contribution of resolved and unresolved MBHBs emitting GWs in the LISA band, computed for each of the previously described combinations of $\beta$ and $z_0$.
Conversely, the right panel shows the same quantity computed for only the unresolved binaries, assuming a detection threshold of $\rho_{\rm th} = 12$.
Both panels also include a solid red curve showing $\OmegaGW(f)$---\textit{total} on the left and background $\OmegaAstro(f)$ on the right---computed for the fiducial \LightSeed~model, to serve as a baseline for comparison across different redshift distributions. 
The fiducial values of $\alpha$, $\M_*$, $\beta$, and $z_0$ used to compute the red curve are listed in Tab.~\ref{tab:pop_model_params}.

The left panel of Fig.~\ref{fig:OmegaAstro_z_dependence} shows that the assumed redshift distribution of the MBHB source population \textit{does} affect the \textit{total} $\OmegaGW(f)$ arising from the superposition of GWs from resolved and unresolved sources in the LISA band.
Specifically, the MBHB population with $\beta = 3$ and $z_0 = 0.75$ (yellow, dot-dashed curve) produces a \textit{total} $\OmegaGW(f)$ signal nearly 10 times larger than that of the fiducial \LightSeed~scenario. 
Conversely, increasing $\beta$, $z_0$, or both, reduces the amplitude of $\OmegaGW(f)$, with $\beta = 10$ and $z_0 = 3$ yielding a signal with half the amplitude of the fiducial \LightSeed~model.
This sensitivity to $\beta$ and $z_0$ stands in contrast to expectations from populations of purely inspiraling binaries, whose predicted $\OmegaGW(f) \propto f^{2/3}$ scaling shows minimal dependence on the redshift distribution of the source population~\cite{Phinney:2001di}.
As a result, our analytical approximation for the \textit{total} $\OmegaGW(f)$ reveals that accounting for the merger of MBHBs (via the $f_{r,{\rm ISCO}}$ cutoff) in the assumed individual source SED [Eq.~\eqref{eq:individual_MBHB_SED}] introduces a stronger dependence on the redshift distribution.

This effect arises from both the greater loudness of nearby sources and the redshifting of GW emission frequencies from individual MBHBs.
In other words, a binary with source-frame chirp mass $\M$ will not only produce a lower-amplitude signal at higher $z$, but will also emit GWs at lower observed frequencies.
Therefore, the fractional contribution from high-energy GW sources (with $\M \gtrsim 10^8\, \Msun$) in each $f$-bin is reduced, as the most massive MBHBs at high redshifts merge below the lowest accessible frequency in the LISA band. 
In contrast, populations dominated by low-redshift sources consist of MBHB systems that contribute to $\OmegaGW(f)$ up to higher observed frequencies, $f_{\rm ISCO} = f_{r,{\rm ISCO}}(1+z)^{-1}$, boosting the overall amplitude of the background.

The right panel of Fig.~\ref{fig:OmegaAstro_z_dependence}, which shows how the unresolved background $\OmegaAstro(f)$ depends on the source redshift distribution, reveals a strikingly different trend. 
Most notably, despite the wide range of redshift distributions considered, the colored curves nearly overlap for  $5\times10^{-5}\ \textrm{Hz} \lesssim f\lesssim 5\times 10^{-2}\ \textrm{Hz}$. This indicates that the application of the individual detection threshold $\rho_{\rm th}$ to isolate the unresolved background from the total $\OmegaGW(f)$ [detailed in Eq.~\eqref{eqn:final_astro_background}] makes the resulting $\OmegaAstro(f)$ largely insensitive to changes in $\beta$ and $z_0$  in the frequency range corresponding to LISA's highest sensitivity.
In contrast, at $f \lesssim 5 \times 10^{-5}$ Hz, the curves separate from the fiducial \LightSeed~$\OmegaAstro(f)$, with the $\{\beta, z_0\} = \{10, 3\}$ distribution producing the loudest signal and $\{\beta, z_0\} = \{3, 0.75\}$ resulting in the weakest.
A similar separation appears at high frequencies, but the order of signal strength is reversed. 
At $f \gtrsim 5 \times 10^{-2}$ Hz, the $\{\beta, z_0\} = \{3, 0.75\}$ distribution produces the strongest $\OmegaAstro(f)$, while $\{\beta, z_0\} = \{10, 3\}$ yields the weakest.

This complex behavior is attributed to a combination of effects: the redshifting of the GW emission of individual MBHBs, source-redshift dependence of the amplitude of emission and how these variations impact the resolvability of MBHBs.
First, given the SNR contours presented in Fig.~\ref{fig:SNR_contour} and the description of the fiducial $\OmegaAstro(f)$ for the \LightSeed\ scenario in Sec.~\ref{sec:astrophysical_background}, it is clear that the regime $5\times10^{-5}\ \textrm{Hz} \lesssim f\lesssim 5\times 10^{-2}\ \textrm{Hz}$ is dominated by an ensemble of low-$\M$, inspiraling MBHBs. 
As a result, in this frequency regime, the dependence of $\OmegaAstro(f)$ on $\beta$ and $z_0$ reduces to that expected from an ensemble of inspiraling sources---the redshift distribution of the ensemble has minimal impact on the amplitude, with $\OmegaAstro(f) \propto f^{2/3}$ remaining unchanged. 
In the low-frequency regime, the $f$-dependence of $\OmegaAstro(f)$ is dominated by the resolvability of the signal from MBHBs with $\M\gtrsim 10^8\Msun$ (see discussion in Sec.~\ref{sec:astrophysical_background}).
Considering the contours in Fig.~\ref{fig:SNR_contour}, a source population dominated by low-redshift MBHBs ensures that individual sources are resolved out to higher masses. 
This again reflects the decreased amplitude of signal observed from high-$z$ sources, as well as the redshifting of their GW emission as discussed above.
Therefore, for populations dominated by high-$z$ sources, the detector is unlikely to resolve the most massive binaries at high redshifts, allowing them to contribute to the background $\OmegaAstro(f)$ and enhance the signal in the low-$f$ regime. 
Finally, in the high-frequency regime, the $f$-dependence of $\OmegaAstro(f)$ is determined by the population of unresolved low-$\M$ binaries with $\M \lesssim 10^4\Msun$ (detailed in Sec.~\ref{sec:astrophysical_background}). 
In this case, the relative signal-strength predicted across the various redshift distribution models is revealed by considering the redshifting of the source-frame $f_{r,{\rm ISCO}}$ for each MBHB; populations dominated by low-$z$ sources present a higher fraction of MBHBs that truncate at higher $f_{\rm ISCO}$ as discussed above.

To conclude, our analysis indicates that the sensitivity of $\OmegaGW(f)$ (\textit{total} or unresolved) to the redshift distribution of the underlying source population is determined by whether the binaries in consideration are merging in band. 
If $\OmegaGW(f)$ can be largely characterized by an ensemble of inspiraling sources, the dependence of the signal on parameters $\beta$ and $z_0$ is nearly negligible. 
However, if the frequency regime in consideration corresponds to signal from merging sources, the amplitude of $\OmegaGW(f)$ is more sensitive to $\beta$ and $z_0$.
Specifically focusing on $\OmegaAstro(f)$ (produced by unresolved MBHBs) we find that the signal is weakly dependent on $\beta$ and $z_0$ in the frequency regime corresponding to LISA's peak sensitivity. 
Although variations in the predicted $\OmegaAstro(f)$ are visible in the low- and high- frequency regimes, these relative changes in signal strength contribute minimally to the SNR. 
Therefore, we expect our forecasted SNRs and parameter errors to be unaffected even by large changes in the assumed redshift distribution of MBHBs.

\section{Forecasts Assuming Flat-Spectrum Cosmological Background}
\label{app:flat_spectrum_forecasts}

\begin{figure}[t]
    \centering
    \includegraphics[width=0.5\textwidth]{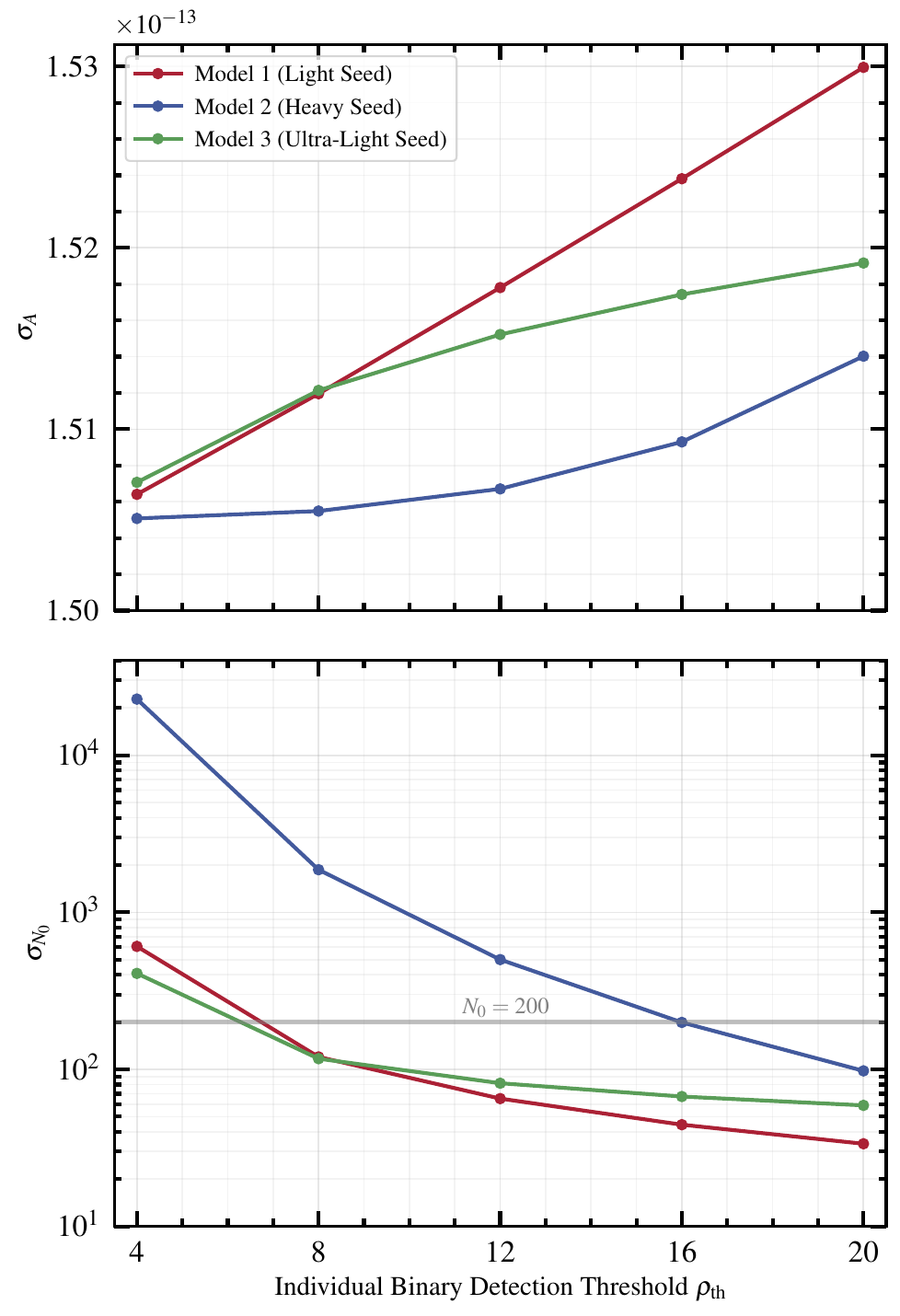} 
    \caption{
    Forecasted $1\sigma$ uncertainties on the amplitude $A$ of a cosmological SGWB (top) and the local merger rate $N_0$ of an astrophysical background (bottom) as functions of the individual binary detection threshold $\rho_{\rm th}$. 
    In this case, we assume a cosmological background characterized by amplitude only (i.e., fixed spectral index $\gamma=0$), jointly analyzed with an astrophysical background for each of the three population models shown in Fig.~\ref{fig:pop_models}. 
    Both uncertainties are computed via information-matrix analysis and are independent of the fiducial value of $A$, as expected. 
    In the bottom panel, the horizontal gray line denotes the fiducial merger rate $N_0 = 200~\mathrm{yr}^{-1}$; uncertainties above this line indicate that the astrophysical background cannot be effectively constrained. 
    Increasing $\rho_{\rm th}$ reduces contamination from individually resolvable binaries, improving the measurement of $N_0$ at the cost of increased uncertainty on $A$.
    }
    \label{fig:sigmaA_sigmaN0_vs_rho_th}
\end{figure}

In this section, we present information-matrix forecasts for the simplified case of a cosmological SGWB with fixed spectral index $\gamma = 0$, motivated by early-Universe models predicting scale-invariant GW backgrounds~\cite{Starobinsky:1979ty,Guzzetti:2016mkm}.

Here, we restrict the parameter space to just the amplitude $A$ of the cosmological background and the local merger rate $N_0$ of the astrophysical background. 
This allows a focused two-parameter information-matrix forecast, which complements the more general three-parameter analysis in Sec.~\ref{sec:Fisher_forecasts}.

\paragraph*{\textbf{Impact of varying $\rho_{\rm th}$.}} 
Figure~\ref{fig:sigmaA_sigmaN0_vs_rho_th} shows the forecasted $1\sigma$ uncertainties on the cosmological amplitude $A$ (top) and the astrophysical merger rate $N_0$ (bottom) as functions of the individual binary detection threshold $\rho_{\rm th}$. 
All results assume a fiducial value of local merger rate $N_0 = 200~\mathrm{yr}^{-1}$. 
As in the three-parameter analysis, increasing $\rho_{\rm th}$ increases the number of unresolved sources, thereby boosting the astrophysical background and improving constraints on $N_0$, while modestly degrading the precision on $A$.

The top panel shows that this degradation in $\sigma_A$ is relatively mild. 
This reflects the fact that detectability is not primarily limited by the amplitude of the astrophysical background, but rather by its spectral similarity to the cosmological signal. 
Even though a higher $\rho_{\rm th}$ results in a louder $\OmegaAstro(f)$, the flat spectral shape ($\gamma=0$) is still distinguishable from the $f^{2/3}$ scaling of the astrophysical component.

The bottom panel confirms that $N_0$ becomes progressively better constrained with higher $\rho_{\rm th}$, consistent with intuition: more unresolved binaries yield a higher SNR background, improving the leverage on $N_0$. 
As anticipated from the background SNRs presented in Tab.~\ref{tab:snr_background_models}, Model~1 (\texttt{Light Seed}) consistently yields the strongest background and hence the smallest $\sigma_{N_0}$ (except at $\rho_{\rm th} = 4$, where Model~3 has slightly better constraints). 
Similarly, the highest $\sigma_A$ values are associated with Model~1, again reflecting the stronger contamination.

\paragraph*{\textbf{Smallest measurable cosmological amplitude.}} 
From this analysis, we find that the smallest detectable cosmological background (at $3\sigma$) corresponds to $A \gtrsim 4.5 \times 10^{-13}$ for most population models and $\rho_{\rm th}$ values. 
A $5\sigma$ detection will require $A \gtrsim 7.5 \times 10^{-13}$. 
While there is mild variation across models and thresholds, these limits are accurate at the order-of-magnitude level.

For comparison, we also performed a baseline forecast assuming only the cosmological background (flat spectrum, and no astrophysical component). 
In that case, we find a forecasted uncertainty $\sigma_A \simeq 4.2 \times 10^{-14}$. 
Comparing this to the joint (astro+cosmo) analysis, we find that the inclusion of the astrophysical foreground increases the minimum detectable amplitude of a cosmological background by a factor of $\sim 3$. 
This underscores the importance of accounting for MBHB confusion noise, even in the flat-spectrum case.

\paragraph*{\textbf{Impact of varying local merger rate $N_0$.}} 
In addition to varying $\rho_{\rm th}$, we also explore how the detectability of the cosmological background depends on the assumed fiducial value of the astrophysical merger rate $N_0$. 
This analysis enables us to evaluate to what extent the robustness of our conclusions depends on the uncertainties in the underlying MBHB population.

\begin{figure}[t]
    \centering
    \includegraphics[width=0.5\textwidth]{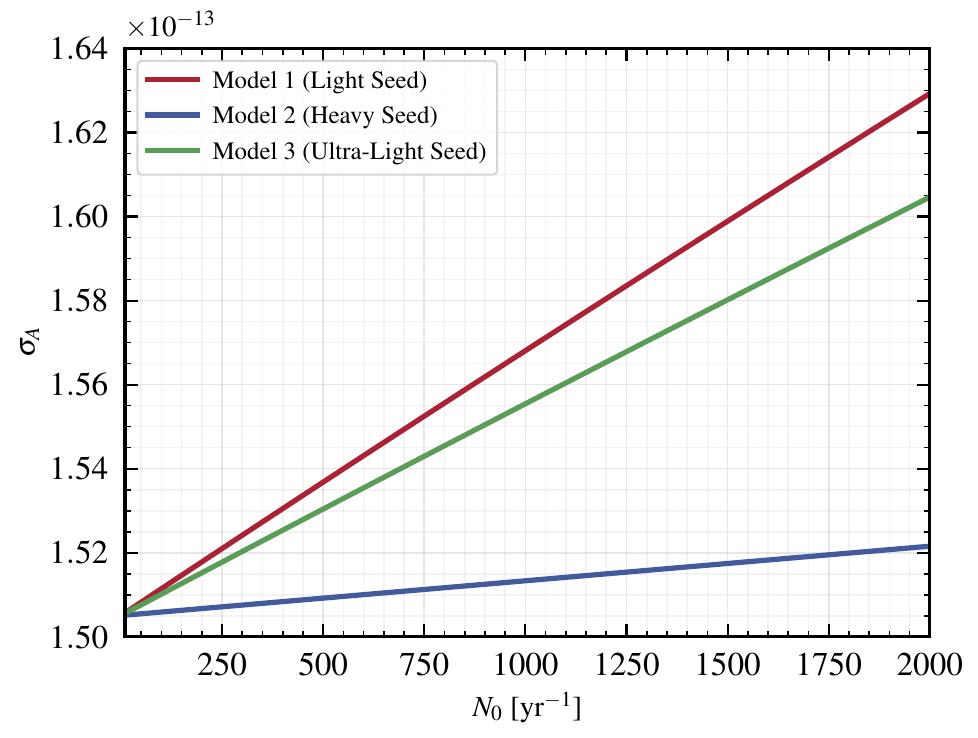} 
    \caption{
    Forecasted $1\sigma$ uncertainty on the amplitude $A$ of a cosmological SGWB as a function of the fiducial local merger rate $N_0$ of the astrophysical background. 
    We assume a cosmological background characterized by amplitude only (i.e., fixed spectral index $\gamma=0$) and perform a joint analysis with an astrophysical background, using the same setup as in Fig.~\ref{fig:sigmaA_sigmaN0_vs_rho_th}. 
    Different curves correspond to the three astrophysical population models defined in Fig.~\ref{fig:pop_models}. 
    Increasing $N_0$ raises the contribution of the unresolved astrophysical background, leading to greater contamination and slightly larger uncertainties in the measurement of $A$. 
    }
    \label{fig:sigmaA_vs_N0}
\end{figure}

Figure~\ref{fig:sigmaA_vs_N0} shows the $1\sigma$ uncertainty on $A$ as a function of the assumed merger rate $N_0$, spanning the range $20~\mathrm{yr}^{-1} \leq N_0 \leq 2000~\mathrm{yr}^{-1}$ (i.e., an order of magnitude above and below our fiducial choice). 
The detection threshold is fixed at $\rho_{\rm th} = 12$ to isolate the effect of $N_0$ alone.

As expected, increasing $N_0$ leads to a higher level of unresolved astrophysical background, which increases the confusion noise and slightly degrades sensitivity to the cosmological signal.
However, the resulting change in $\sigma_A$ is remarkably modest: across two orders of magnitude in $N_0$, the uncertainty increases by less than 10\%.
This insensitivity indicates that our conclusions about detectability remain robust even if the true MBHB merger rate is an order of magnitude smaller than our fiducial assumption. 

The degree of degradation varies by model, consistent with their background SNRs reported in Tab.~\ref{tab:snr_background_models}. 
Model~1 (\texttt{Light Seed}) yields the highest $\sigma_A$ due to its stronger background, followed by Model~3 (\texttt{Ultra-Light Seed}), while Model~2 (\texttt{Heavy Seed}) exhibits the smallest degradation, reflecting the fact that most binaries in this scenario are resolvable, and therefore, do not contribute to the confusion-noise-limited background.

We emphasize again that the detectability of $A$ depends more on the spectral similarity between the cosmological and astrophysical components than on the absolute level of the astrophysical background. 
Thus, even substantial variation in $N_0$ does not significantly affect $\sigma_A$ unless it also alters the shape of the unresolved component. 
It is important to note, however, that increasing $N_0$ non-trivially affects the detectability of individual sources. 
In particular, if the MBHB merger rate in the LISA band is very high, our resolved binary subtraction procedure (described in Sec.~\ref{sec:astrophysical_background}) likely overestimates the resolvability of individual events, leading to an underestimated prediction for $\OmegaAstro(f)$. 
Therefore, the forecasts shown in Fig.~\ref{fig:sigmaA_vs_N0} can be considered a lower bound on $\sigma_A$, especially at high $N_0$.
Moreover, if $N_0$ can be constrained independently using resolved MBHBs, the uncertainty on $A$ can be reduced.

\bibliography{lisa_bkgrd}

\end{document}